\documentclass{natureJLH}

\usepackage{graphicx}
\usepackage[none]{hyphenat}
\usepackage{tikz}
\usepackage[normalem]{ulem} 
\usepackage{lineno}
\usepackage{amssymb}
\usepackage{amsmath}
\usepackage{caption}
\usepackage{booktabs,multirow}
\usepackage{multirow}
\usepackage{pdflscape}
\usepackage{xcolor}
\usepackage{hyperref}

\newcommand{\araa}{Annu. Rev. Astron. Astrophys.}   
\newcommand{\aj}{Astron. J.}   
\newcommand{\apj}{Astrophys. J.}   
\newcommand{\apjl}{Astrophys. J. Lett.}   
\newcommand{\apjs}{Astrophys. J. Suppl. Ser.}   
\newcommand{\apss}{Astrophys. Space Sci.}   
\newcommand{\aap}{Astron. Astrophys.}   
\newcommand{\aapr}{Astron. Astrophys. Rev.}   
\newcommand{\memsai}{Mem. Soc. Astron. Italiana}   
\newcommand{\mnras}{Mon. Not. R. Astron. Soc.}   
\newcommand{\nat}{Nature} 
\newcommand{\prl}{Phys. Rev. Lett.}   
\newcommand{\pasa}{Publ. Astron. Soc. Aust.}   
\newcommand{\pasp}{Publ. Astron. Soc. Pac.}   

\newcommand{\msun}{\mbox{$M_{\odot}$}}
\newcommand{\rsun}{\mbox{$R_{\odot}$}}

\title{A dynamically discovered and characterized non-accreting neutron star - M dwarf binary candidate}

\author{Tuan Yi$^{1}$, Wei-Min~Gu$^{*1}$, Zhi-Xiang~Zhang$^{1}$, 
Ling-Lin~Zheng$^{1}$, Mouyuan~Sun$^{*1}$, Junfeng~Wang$^{1}$, 
Zhongrui~Bai$^{2}$, Pei~Wang$^{2}$, Jianfeng~Wu$^{1}$, 
Yu~Bai$^{2}$, Song~Wang$^{2}$, Haotong~Zhang$^{2}$, 
Yize~Dong$^{3}$, Yong~Shao$^{4}$, Xiang-Dong~Li$^{4}$,
Jia~Zhang$^{5}$, Yang~Huang$^{6}$, Fan~Yang$^{7}$,
Qingzheng~Yu$^{1}$, Hui-Jun Mu$^{8}$, Jin-Bo~Fu$^{1}$,
Senyu~Qi$^{1}$, Jing~Guo$^{1}$, Xuan~Fang$^{2}$,
Chuanjie~Zheng$^{2,9}$, Chun-Qian~Li$^{2}$, Jian-Rong~Shi$^{2,9}$,
Huanyang Chen$^{10}$, and Jifeng~Liu$^{*2,9}$. }

\begin{document}

\maketitle

\begin{affiliations}

\item Department of Astronomy, Xiamen University, Xiamen, Fujian 361005, P. R. China
\item National Astronomical Observatories, Chinese Academy of Sciences, Beijing 100101, P. R. China
\item Department of Physics and Astronomy, University of California, 1 Shields Avenue, Davis, CA 95616-5270, USA
\item Department of Astronomy, Nanjing University, Nanjing 210046, P. R. China
\item Yunnan Observatories, Chinese Academy of Sciences, Kunming 650216, P. R. China
\item South-Western Institute for Astronomy Research, Yunnan University,
Kunming 650500, P. R. China
\item Department of Astronomy, Beijing Normal University, Beijing 100875, P. R. China
\item International Laboratory for Quantum Functional Materials of Henan, and School of Physics and Microelectronics,
Zhengzhou University, Zhengzhou, Henan 450001, P. R. China
\item School of Astronomy and Space Science, University of Chinese Academy of Sciences, Beijing 100049, P. R. China
\item Department of Physics, Xiamen University, Xiamen, Fujian 361005, P. R. China

\end{affiliations}

\begin{abstract}

Optical time-domain surveys can unveil and characterize exciting but less-explored non-accreting and/or non-beaming neutron stars (NS) in binaries. Here we report the discovery of such a NS candidate using the LAMOST spectroscopic survey. 
The candidate, designated LAMOST J112306.9+400736 (hereafter J1123), 
is in a single-lined spectroscopic binary containing an optically visible M star. 
The star's large radial velocity variation and ellipsoidal variations 
indicate a relatively massive unseen companion. 
Utilizing follow-up spectroscopy from the Palomar 200-inch telescope and high-precision photometry from TESS, 
we measure a companion mass of $1.24_{-0.03}^{+0.03}~M_{\odot}$. 
Main-sequence stars with this mass are ruled out, 
leaving a NS or a massive white dwarf (WD). 
Although a massive WD cannot be ruled out, 
the lack of UV excess radiation from the companion supports the NS hypothesis. 
Deep radio observations with FAST yielded no detections of either pulsed or persistent emission. 
J1123 is not detected in numerous X-ray and gamma-ray surveys. 
These non-detections suggest that the NS candidate is not presently accreting 
and pulsing.
Our work exemplifies the capability of discovering compact objects in non-accreting close binaries 
by synergizing the optical time-domain spectroscopy and high-cadence photometry.

\end{abstract}

The demographic and physical properties 
(e.g., birth rates and mass distributions\cite{Ozel2012}) 
of NSs hold essential information about 
the stellar evolution and chemical enrichment history of our Galaxy.
NSs are typically discovered with different manifestations\cite{Keane2008} 
in the electromagnetic window, 
e.g., rapidly rotating and highly magnetized radio pulsars\cite{Lorimer2008}, 
accreting NS X-ray binaries, gamma-ray pulsars\cite{Abdo2010}, 
and nearby isolated 
thermally emitting NSs\cite{Haberl2007}. 
In recent years,
coalescing NSs that produce gravitational waves\cite{Abbott2017}
have also been detected by LIGO and Virgo.
However, samples of currently discovered NSs 
are far from being complete and representative.
Since radio, X-ray, or gamma-ray observations discover NSs 
only if they are beaming towards us or the NS is accreting material from its companion,
the population of non-accreting and/or radio quiet NSs\cite{Caraveo1996} 
remains largely undetected.

In the optical window, large spectroscopic surveys like 
LAMOST\cite{Cui2012,Zhao2012}
(Large Sky Area Multi-Object Fiber Spectroscopic Telescope)
have been monitoring millions of stars 
and archiving the largest spectroscopic database.
By mining the wealth of large spectroscopic databases,
one can implement the radial velocity (RV) method to discover systems
that harbor a hidden compact object in orbiting with a luminous stellar companion\cite{Trimble1969},
e.g., stellar-mass black holes (BH).
A handful of notable systems have been discovered recently, 
such as MWC 656\cite{Casares2014}, LB-1\cite{Liu2019}, and J05215658+4359220\cite{Thompson2019}. 
These BHs reside in wide binaries with an orbital period typically $\gtrsim$ 50 days.
By synergizing multi-epoch spectroscopic and photometric surveys, 
one can dynamically measure their masses.
In principle, searching for NSs using this methodology would be feasible.
Surprisingly, only two potential candidates 
have been discovered by the dynamical method to date, 
i.e., J05215658+4359220\cite{Thompson2019} and V723 Mon\cite{Jayasinghe2021}.
While each system might contain a massive NS, 
a non-interacting, mass-gap\cite{Bailyn1998} BH
($\sim$ 3 $M_{\odot}$) was considered more likely.

\begin{figure}
\centering
\includegraphics[width=0.8\linewidth]{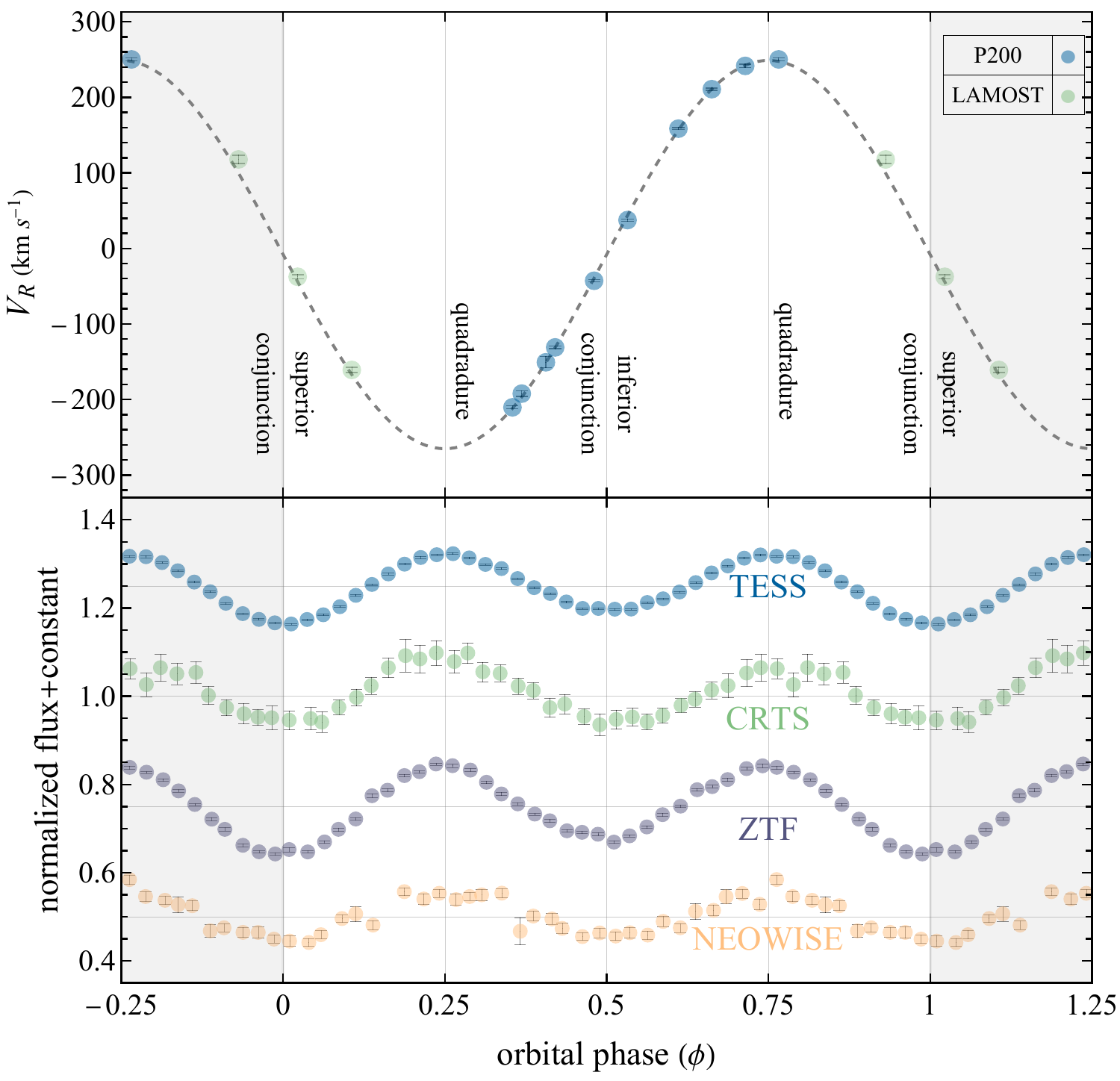}
\caption{The phase-folded RV curve and the multi-band light curves for J1123.
Upper panel: the RV curve. 
{The green and blue points represent the exposures of LAMOST and P200, respectively.  
Lower panel: the phase-folded and smoothed light curves from
TESS (blue), CRTS (green), ZTF r-band (purple), and NEOWISE W1-band (orange).}
All data are phase-folded with the ephemeris given in Equation~(\ref{eq:ephem}),
and smoothed by grouping the data into 40 equally spaced phase bins.
The average within each bin is adopted. 
Data points are styled semi-transparent so the error bars 
(representing the $1$-sigma uncertainties) can be seen
(zoom in to find black fences representing the error bars inside).
The superior conjunction ($\phi$ $=$ 0 or 1), inferior conjunction ($\phi$ $=$ 0.5), 
and quadrature phases ($\phi$ $=$ 0.25 and 0.75) of the M dwarf are labeled 
and indicated by vertical grey lines. 
Horizontal grey lines in the lower panel indicate the mean of the normalized and vertically shifted
light curve for each band.
Points in shallow grey regions are duplicated ones for clarity.
The RV curve and the light curves demonstrate the periodic 
\textit{Doppler} shifts and ellipsoidal modulation effects
caused by J1123's invisible companion's strong gravitational pull and tidal force.
}
\label{fig:rvlc}
\end{figure}

Here we report the discovery of a close binary system with a non-accreting NS candidate.
This system, designated LAMOST J112306.9+400736 (hereafter J1123), 
is a single-lined spectroscopic binary with an early-type M dwarf at a distance of 318 pc. 
On February 22, 2015, the M-star showed a radial velocity change of $\sim$ 270 km~s$^{-1}$ within $\sim$ 68 minutes. Such a large and rapid velocity shift and the characteristic ellipsoidal variations\cite{Morris1993} showing on multi-band light curves (Figure~\ref{fig:rvlc}; lower panels) 
indicate a relatively massive unseen companion. 
Archival data and follow-up observations confirm the presence of a compact companion, which is either a NS or a massive and cold white dwarf (WD).


\section{Results} 
Following the discovery of J1123, 
we took follow-up spectroscopy 
by using the Double Spectrograph (DBSP) mounted on 
Palomar's 200-inch telescope (P200), 
to obtain ten more exposures 
(Figure~\ref{fig:rvlc}; upper panel).
Utilizing the multi-epoch spectroscopy and photometry, 
we measure the dynamical mass of the unseen companion to be
$1.24_{-0.03}^{+0.03}~M_{\odot}$.
Because normal stars with similar masses would easily
outshine the visible M dwarf, 
we conclude that the unseen component must be a compact object.
Supplementary Table 1 
summarizes the stellar and binary parameters of J1123.

\subsection{The orbital ephemeris, radial velocity curve, and the mass function} 

The orbital ephemeris of J1123 is:
\begin{equation}\label{eq:ephem}
T(\phi = 0) = 
2\,453\,734.909\,32(31) \mathrm{HJD} +
0.273\,835\,44(18) \times N \ ,
\end{equation}
where the first term in the right hand is an epoch of the superior conjunction
(denoted as $T_{0}$, i.e., corresponding to phase $\phi$ $=$ 0 where the M dwarf 
is farthest away from the observer), HJD is the Heliocentric Julian Date, 
and the second term is the orbital period times the number of orbital cycles $N$ (see Methods).
Last two digits inside the parenthesis indicate the one-sigma uncertainties.

We measure the RVs by cross-correlating the spectra 
to a library of empirical spectra\cite{Kesseli2017} 
(see Methods). 
Both RVs and photometric data are phase-folded 
according to Equation~(\ref{eq:ephem}).
We assume that the orbit is circularized 
and use a sinusoidal curve 
$V_\mathrm{R} = - K_{1} \sin (2\pi(t-T_{0})/P_\mathrm{orb}) + \gamma$ 
to fit the RV data (Figure~\ref{fig:rvlc}), 
where $K_{1}$ is the RV curve semi-amplitude,
$t$ is the mid-time of each exposure, 
$P_{\rm orb}$ is the orbital period,
and $\gamma$ is the systemic velocity.
The fitting results in $K_{1}$ $=$ 257 $\pm$ 2 km~s$^{-1}$
and $\gamma$ $=$ $-$8 $\pm$ 2 km~s$^{-1}$. 
Hence the mass function (lower mass limit) for the invisible companion is: 
\begin{equation}\label{eq:fm}
f(M_{\rm 2}) = \frac{M_{2}^{3} \sin^{3} (i)}{(M_{1}+M_{2})^2}
= \frac{K_{1}^{3} P_{\rm orb}}{2 \pi G} = 
0.48\pm 0.01~M_{\odot} \ ,
\end{equation}
where $M_{1}$ and $M_{2}$ are the masses of the visible and invisible components, 
respectively, $i$ is the orbital inclination, 
and $G$ is the gravitational constant.

\subsection{The stellar properties of the M dwarf}

The visible star has a spectral type of M1, 
featuring the TiO molecular absorption bands of a typical M dwarf.
Emission lines of
\mbox{Ca II H\&K} and Balmer series (H$\alpha$ -- H$\gamma$) 
are presented in all of J1123's spectra. 
We found that these emission lines are co-moving with the M dwarf,
suggesting that they originate from the hot chromosphere of the 
M dwarf (see Methods).

\begin{figure}
\centering
\includegraphics[width=1.0\linewidth]{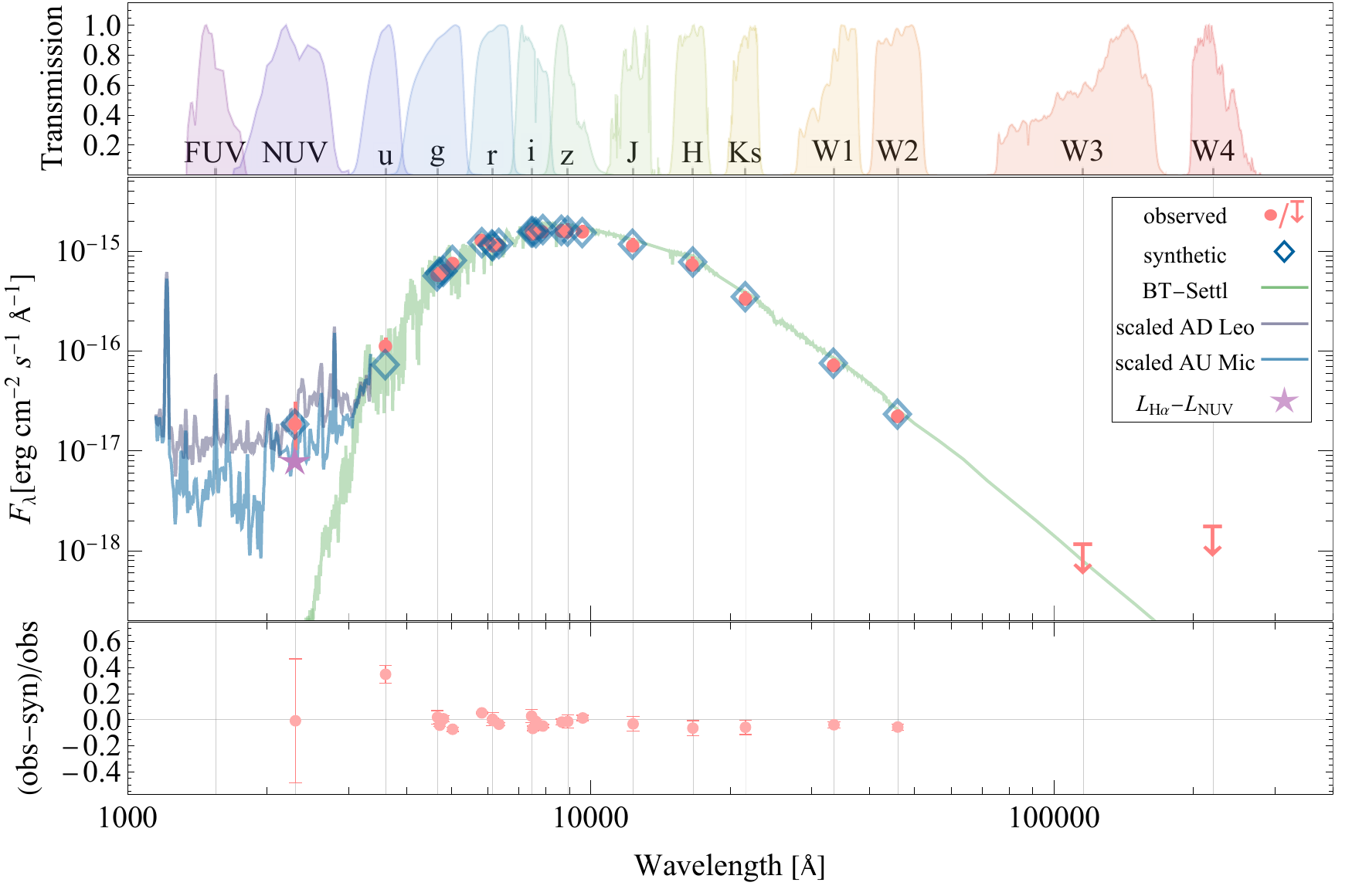}
\caption{The broadband SED of J1123. 
Pink points are the observed flux 
from the GALEX NUV band, SDSS $ugriz$ bands, ZTF $gri$ bands, 
{\it Gaia} EDR3 BP-, RP- and G-bands, PanSTARRS $grizy$, 
2MASS J H Ks bands, and AllWISE W1-W4 bands 
(bands for PanSTARRS, ZTF, and {\it Gaia} are not labeled 
to avoid overcrowding).
Error bars represent the $1$-sigma uncertainties and downward arrows indicate upper limits for W3 and W4 bands.
The green curve is the best-fit BT-Settl model spectrum for the M dwarf
and blue hollow diamonds are the synthetic photometry 
by integrating the model SED over the transmission curve of each filter.
Transmission curves for the filters of GALEX, SDSS, 2MASS, and WISE 
are displayed on the top panel and normalised at the maximum.
To demonstrate that the observed UV excess is fully consistent with being 
produced by the chromospheric activities of the visible M dwarf,
we scale on top of the SED the average UV spectrum of AD Leo (purple line), 
and of AU Mic (blue line). 
The mean synthetic NUV flux of these two M stars well matches the observation.
Showing in the lower panel are the relative fit residuals,
which is defined as (observed $-$ synthetic) $/$ observed. 
A predicted NUV flux for a sample of similar active M dwarfs\cite{Jones2016}
is also shown by the purple star (see text for the estimation by an empirical 
$L_\mathrm{H\alpha}$-$L_\mathrm{NUV}$ relation\cite{Jones2016}).
} 
\label{fig:sed}
\end{figure}

We fit the broadband SED (Figure~\ref{fig:sed}) of J1123 using its archival photometry
and utilizing the distance $D = 318 \pm 4$ pc derived from the parallax 
$\bar \omega$ $=$ 3.147 $\pm$ 0.038~mas ({\it Gaia} EDR3\cite{Gaia2021}).
The best-fit model SED has an effective temperature 
$T_\mathrm{eff,1}$ $=$ 3769$_{-9}^{+8}$ K,
a radius $R_{1}$ $=$ 0.63 $\pm$ 0.01 $R_{\odot}$,
and a bolometric luminosity
$L_\mathrm{bol}$ $=$ 4 $\pi R_{1}^{2} \sigma T_\mathrm{eff,1}^{4}$ 
$=$ 0.071 $\pm$ 0.002$L_{\odot}$.
We measure a stellar mass 
$M_{1}$ $=$ 0.61 $\pm$ 0.02 $M_{\odot}$ 
of the M dwarf by fitting the stellar model isochrones (see Methods).

\subsection{The orbital solution: unveiling the nature of the invisible}

\begin{figure}
\centering
\includegraphics[width=0.8\linewidth]{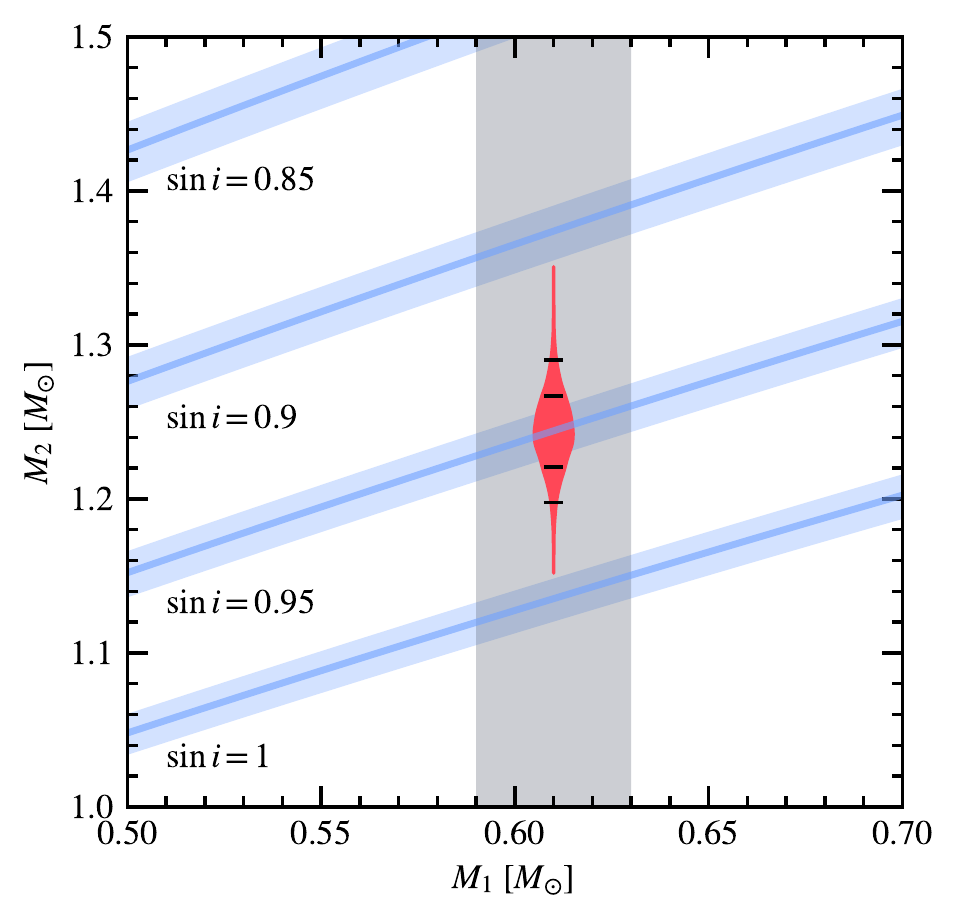}
\caption{The mass of the visible star ($M_1$) versus that of the hidden compact 
object ($M_2$). The violin width indicates the probability density distribution of $M_2$, 
which is obtained by fitting the ellipsoidal light-curve model to the \texttt{TESS} 
observations and RV variations for fixed $M_1$ and $R_1$. The four black short lines 
demonstrate the $1$-sigma and $2$-sigma ranges. The four solid blue curves correspond to 
the constraints from the mass function (Equation~\ref{eq:fm}) for $\sin i$ $=$ 1, 0.95, 
0.9, and 0.85, respectively; the blue-shaded regions represent the $1\sigma$ uncertainties. 
The grey-shaded regions illustrate the $1$-sigma uncertainties of $M_1$. }
\label{fig:m1m2}
\end{figure}

Equipped with all the observed and derived properties of the visible M dwarf, 
we model the RV curve and the ellipsoidal light curve 
using the \texttt{PHOEBE} software and solve: 
the orbital inclination $i$, the mass ratio $q\, (\equiv M_{2}/M_{1})$, 
and hence the compact object's mass $M_{2}$.
By fixing $M_1 = 0.61\ M_{\odot}$ and $R_1 = 0.63\ R_{\odot}$, we obtain 
the following orbital solution (for a complete discussion, see Methods): 
$i = 73^{+1.8}_{-1.5}$ degree and $M_{2} = 1.24_{-0.03}^{+0.03}~M_{\odot}$ 
(Figure~\ref{fig:m1m2}).
The measured mass suggests that the compact object is either a massive WD or a NS.

\begin{figure}
\centering
\includegraphics[width=1.0\linewidth]{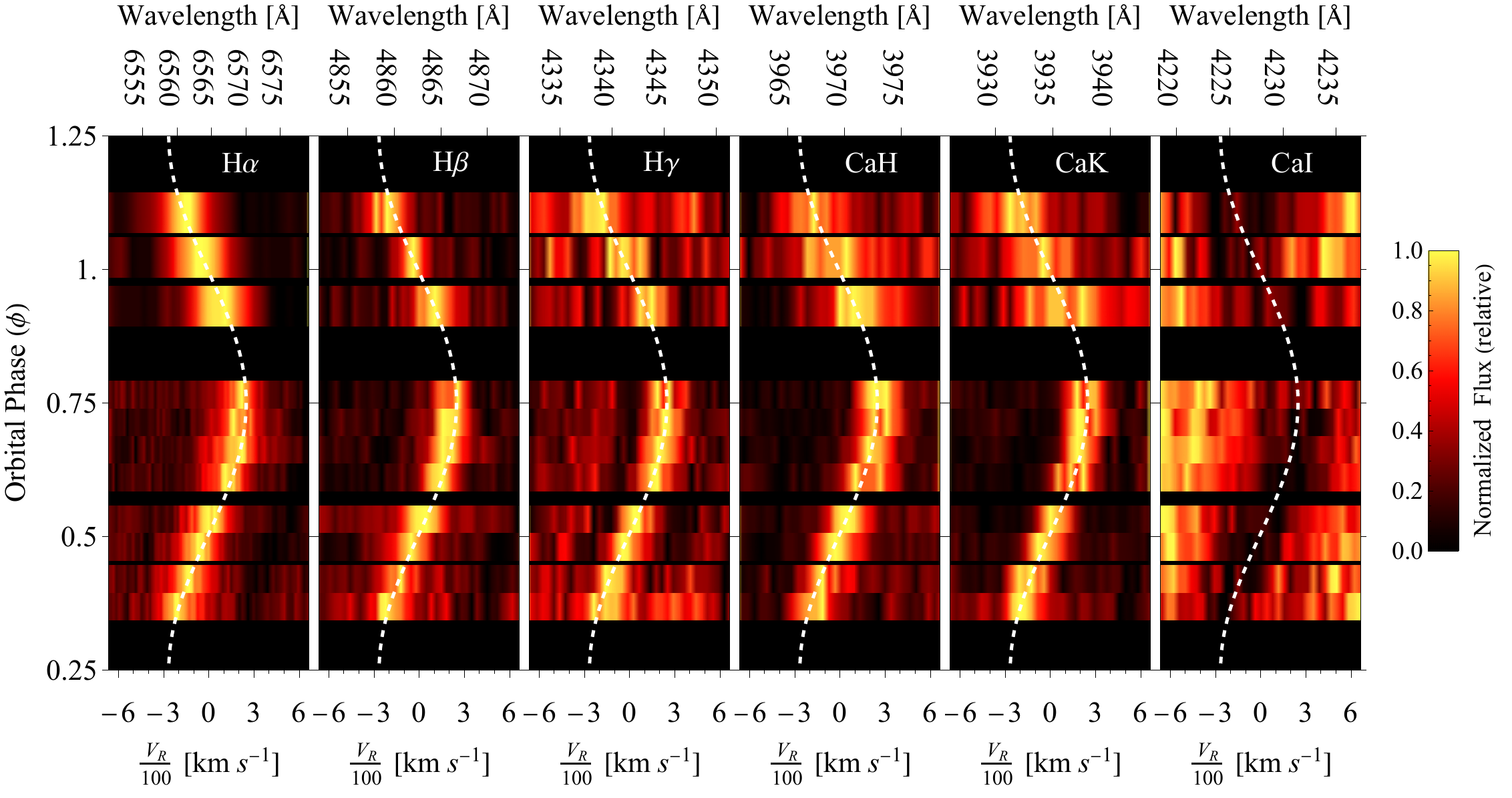}
\caption{Phase-stacked Balmer (H$\alpha$ -- H$\gamma$) emission lines, 
and \mbox{Ca II H\&K} emission lines. 
The white dashed curve threading through the centers of each line 
represent the best-fit RV curve from fitting the photospheric
absorption lines of the M dwarf 
(i.e., using CCF technique with all emission lines and telluric absorption lines masked).
For comparison, the variation of the Ca I 4228\AA{} absorption line is shown in the rightmost panel.
It is evident that the emission lines 
are co-moving with the M dwarf,
which suggests that they originate from the chromosphere of the M dwarf. 
Notes (a): two P200 exposures taken on June 2020 are not plotted to avoid overlapping;
(b): the exposure time is 30~mins for LAMOST and 20~mins for P200.}
\label{fig:balmer}
\end{figure}

We present evidence against a WD.
From the spectroscopic point of view, there are no broad absorption features of a WD 
detected in the vicinity of the Balmer lines, and the P200 spectra are well fitted with the M dwarf template (see Methods).
From the photometric point of view, 
the NUV flux reported by GALEX's All-sky Imaging Survey is well consistent 
with the M dwarf's chromospheric activities. 
Indeed, common
indicators of chromospheric activities\cite{Linsky2017}
like the \mbox{Ca II H\&K} and H$\alpha$ emission lines
are presented in all of the J1123's spectra (Figure~\ref{fig:balmer}).
We calculate the chromospheric NUV luminosity of J1123 according to 
the empirical $L_\mathrm{H\alpha}$ - $L_\mathrm{NUV}$ relation\cite{Jones2016} 
(see Methods) and the observed H$\alpha$ luminosity. We find that the 
chromospheric NUV luminosity can fully account for the NUV observation. 
As a demonstration, we present two properly scaled chromospheric 
UV spectra to the SED: AD Leo (GJ 388), a very active M3.5 dwarf, and the AU Mic, 
a less active M0 dwarf. 
The two scaled spectra can fully fit the observed UV emission in J1123
(Figure~\ref{fig:sed}),
i.e., no WD-contributed UV emission is presented in the SED.
Thus we believe that J1123 is more likely to host a NS.

\subsection{FAST observation of J1123}

We conducted a targeted deep observation of J1123 
using the Five-hundred-meter Aperture Spherical radio Telescope (FAST) on Sep 7, 2021. 
Searches for both persistent radio pulsations and single pulses were performed 
for the 50 minutes integration time, 
but resulted in non-detections with the two types of searches at L-band (1.05 -- 1.45 GHz). 
We measured the amount of pulsed flux above the calibrated baseline noise level, 
giving the 5$\sigma$ upper limit of flux density measurement of 8 $\pm$ 2 $\mu$Jy 
(assuming a pulse duty cycle of 0.1 -- 0.5) for persistent radio pulsations, 
and the stringent upper limit of pulsed radio emission is $\sim$0.012 Jy-ms 
assuming a 1 ms wide burst in terms of integrated flux (see Methods).

Our FAST observation took place when the M dwarf 
was at the orbital phase $\phi$ $=$ 0.24 -- 0.37. 
Hence, radio eclipses due to the M dwarf are less likely.
We stress that many spider systems\cite{Roberts2013,Burdge2022} 
are known to eclipse for extended periods,
due to the scattering of pulses by the stellar companion's wind 
or the swept-back tail of materials\cite{Polzin2020}. 
For instance, pulsar PSR J1748-2446ad has an eclipse fraction of 
around 40\% of the orbital period\cite{Hessels2006}. 
In addition, scintillation effects due to the nearby interstellar medium can also block the radio pulsations.
Thus the non-detection of pulsar signal in a single observation 
towards J1123 is not guaranteed to rule out a pulsar.

\subsection{A non-accreting nearby NS in orbiting with an active M dwarf}

J1123 is a non-accreting system. As discussed, 
the Balmer lines originate from the M dwarf.
We do not see any emission line that moves in anti-phase to the M dwarf's 
motion if it comes from the compact object's side.
A systematic search of the archival ROSAT, \textit{Chandra}, 
\textit{XMM-Newton}, and eROSITA X-ray observations reveals no 
detection.
The nearest ROSAT X-ray source, 2RXS J112241.9+4010.43, 
is about 5.7 arcmin away, thus it is not the X-ray counterpart of J1123.

Since gamma-ray beams are normally wider than radio beams, 
a gamma-ray pulsar\cite{Clark2021} should be easier to detect than a radio pulsar.
However, there is no gamma-ray source associated with J1123. 
As reported by Fermi's 4FGL catalogue\cite{Fermi2020}, 
the nearest source 4FGL J1122.2+3926 is about 42 arcmin 
(or 0.7 deg) away whose localisation is about 0.2 deg at a 95\% confidence level, 
thus the Fermi source cannot be the gamma-ray counterpart of J1123. 
The close distance and the lack of gamma-ray association\cite{Abdo2010} 
suggest that J1123 is different from pulsars like redbacks.
In other words, J1123 is potentially a non-beaming NS.

\begin{figure}
\centering
\includegraphics[width=0.7\linewidth]{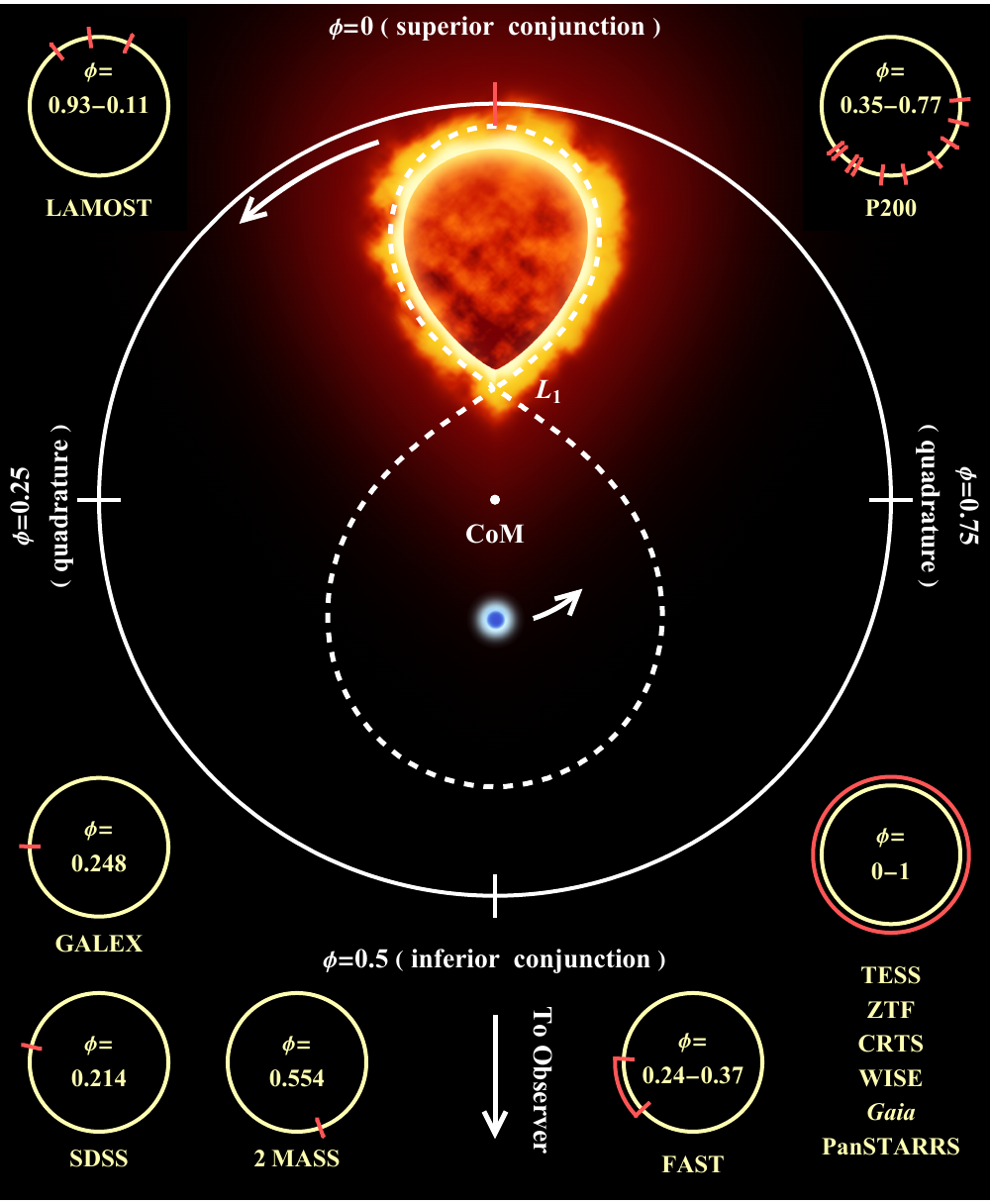}
\caption{The viewing wheel of J1123 that summarises our
best understanding of the system. J1123's only visible component,
an early-type M dwarf is gravitationally pulled
by its invisible companion, a candidate NS (blue object).
Two components are in orbiting with each other with an 
orbital period of $\sim$6.6 hours.
An artistic impression of J1123's active chromosphere is depicted
(bright glowing region above the surface of the star),
which induces substantial UV emissions as well as the emission lines.
No signature of Roche lobe (white dashed line)
overflow between two components suggests that
the system is a non-accreting binary.
Showing in corners, small yellow circles are copies of the orbit, 
and red marks denote the positions (phases) of the M star 
at which a specific observation was conducted. 
}
\label{fig:wheel}
\end{figure}

Figure~\ref{fig:wheel} summarises our current best understanding of J1123: a 
non-accreting NS candidate in orbiting with an early-type M dwarf, and the M dwarf 
is active and exhibiting substantial chromospheric UV emission. Besides, J1123 is 
located only $318 \pm 4$~pc away from the Solar System, making it one of the 
nearest NS candidates (twice the distance compared to the nearest millisecond pulsar 
PSR J0437-4715\cite{Johnston1993}; 157 $\pm$ 2.7pc).

\section{Discussion}

J1123 could be a progenitor of low-mass X-ray binaries (LMXB).
The Roche lobe filling factor of the M dwarf is 
$f \equiv R_\mathrm{1} / R_\mathrm{L1} \simeq 0.9$,
where $R_\mathrm{L1}$ is the Roche lobe radius.
According to the magnetic braking mechanism\cite{Verbunt1981}, 
the angular momentum torque\cite{Rappaport1983}
$\dot{J}_{\rm orb} = -3.8 \times 10^{-30} M_{1} R_{1}^{4} \omega^{3} ~\mathrm{dyn~cm}$ 
($\omega$ is the angular velocity of the system). 
The system's present angular momentum is
$J_{\rm orb} = (M_{1}M_{2})/(M_{1}+M_{2}) a^{2} \omega$.
Hence, we estimated a timescale $(J_{\rm fill} - J_{\rm orb}) / \dot{J}_{\rm orb} \sim 2 \times 10^{7}\, \mathrm{yr}$ 
for the system to become an LMXB, where $J_{\rm fill}$ is the angular momentum
when the M star eventually fills the Roche lobe ($f = 1$). 

Binary population synthesis calculation shows that the birthrate of pre-LMXBs
is $\sim$ $10^{-5} \, \mathrm{yr}^{-1}$ and around $10^{4}$ -- $10^{5}$ LMXBs exist 
in our Galaxy\cite{Shao2015}. 
It can be estimated that there are about 
$2 \times 10^{7}\, \mathrm{yr}$ $\times$ $10^{-5} \, \mathrm{yr}^{-1}$ $ = 200$
systems like J1123 and hence the ratio of the number of J1123-like systems to the number of LMXBs is roughly 1 : 100.  
In some transient LMXBs,
the atmosphere of the donor undergoes 
alternating expansion and contraction subject to the irradiation of the NS. 
Thus J1123 might also be a transient LMXB, 
only that the system is currently in a contraction phase
and manifesting a quiescent state. 

As mentioned, the possibility that a WD resides in J1123 
cannot be fully ruled out.
By analysing the NUV flux and P200 blue arm spectrum, 
we argue that, if the compact object is a massive WD, 
the WD must be cold ($\lesssim 10^4$ K; see Methods).
In this case, the most analogous systems would be 
white dwarf-main sequence (WDMS) binaries 
in post common envelope systems. 
J1123 however, distinguishes from all the WDMSs 
because the M dwarf is absolutely dominant on the spectra.
Thus it can be easily missed by conventional WD surveys.
Only the dynamical method supported by time-domain optical surveys
can uncover it from the hidden side.

Precise astrometry by \textit{Gaia}
is also capable of finding hidden BHs, NSs, WDs, 
even brown dwarf companions, 
yet it is limited to discover wide binaries 
with an orbital period from a few weeks up to years.  
In contrast, the RV method is capable of 
discovering binaries that host compact objects regardless of 
the orbital separations.   
Our work exemplifies the methodology 
by synergizing the RV curve and ellipsoidal light curves
to discover and measure non-accreting and/or non-beaming NSs in close binaries,
which is a majority, but under-explored population.


\clearpage

\clearpage

{\flushleft

\clearpage
\begin{addendum}

\item[Data availability]

The LAMOST low-resolution spectra can be queried at 
\url{https://nadc.china-vo.org/data/data/sedr5/f?&locale=en}. 
Raw TESS data are available from the MAST portal: \url{https://archive.stsci.edu/missions-and-data/tess}. 
The data can also be obtained from the corresponding author upon reasonable request.

\item[Code availability]

The Lomb-Scargle module is publicly available on \url{https://github.com/TuanYi/LombScargle}. 
Reasonable requests for other materials and codes should be addressed to 
Wei-Min Gu (guwm@xmu.edu.cn).

\item[Acknowledgements]
We thank anonymous referees for providing numerous constructive suggestions
that improved the quality of this paper.
We thank Ya-Juan Lei, Daniel Price, Alexander Heger, Cui-Ying Song, 
Di Li, Xiao-Dian Chen, Wei-Kai Zong, Xiang-Gao Wang, 
Qiao-Ya Wu, Jun-Hui Liu, Xiao-Dan Fu, Yuan-Pei Yang, 
Xiao-Wei Liu, Yi-Han Song, and Tom Marsh for beneficial discussions. 
W.M.G. acknowledges support from 
the National Key R\&D Program of China under grant 2021YFA1600401,
and the National Natural Science Foundation of China (NSFC) under grants 11925301 and 12033006.
M.Y.S. acknowledges support from NSFC under grant 11973002.
J.F.L. acknowledges support from NSFC
under grants 11988101 and 11933004.
Z.X.Z. acknowledges
support from NSFC under grant 12103041.
J.F.Wang acknowledges support from NSFC under grant U1831205.
J.F.Wu acknowledges support from NSFC under grant U1938105.
X.D.L. acknowledges support from NSFC under grants 12041301 and 12121003.
P.W. acknowledges support from NSFC under grant U2031117, the Youth Innovation Promotion Association CAS (id. 2021055), CAS Project for Young Scientists in Basic Research (grant YSBR-006) and the Cultivation Project for FAST Scientific Payoff and Research Achievement of CAMS-CAS.
J.R.S. acknowledges support from NSFC under grant 12090044.
J.Z. acknowledges support from NSFC under grant 11933008.
H.J.M. acknowledges support from NSFC under grant 12103047.
T.Y. acknowledges support
from the China Postdoctoral Science Foundation under grant 2021M702742.
Guoshoujing Telescope (the Large Sky Area Multi-Object Fiber Spectroscopic Telescope LAMOST) is a National Major Scientific Project built by the Chinese Academy of Sciences. Funding for the project has been provided by the National Development and Reform Commission. LAMOST is operated and managed by the National Astronomical Observatories, Chinese Academy of Sciences.
This research uses data obtained through the Telescope Access Program (TAP), 
which has been funded by the TAP member institutes.
This work made use of the data from FAST (Five-hundred-meter Aperture Spherical
radio Telescope). FAST is a Chinese national mega-science facility, operated by National Astronomical Observatories,
Chinese Academy of Sciences.
Funding for the TESS mission is provided by NASA’s Science Mission directorate.
This research made use of \texttt{Lightkurve}, a Python package for Kepler and TESS data analysis (Lightkurve Collaboration, 2018\cite{Lightkurve2018}).
Based on observations obtained with the Samuel Oschin 48-inch Telescope at the Palomar Observatory as part of the Zwicky Transient Facility project. ZTF is supported by the National Science Foundation under Grant No. AST-1440341 and a collaboration including Caltech, IPAC, the Weizmann Institute for Science, the Oskar Klein Center at Stockholm University, the University of Maryland, the University of Washington, Deutsches Elektronen-Synchrotron and Humboldt University, Los Alamos National Laboratories, the TANGO Consortium of Taiwan, the University of Wisconsin at Milwaukee, and Lawrence Berkeley National Laboratories. Operations are conducted by COO, IPAC, and UW.

\item[Author contributions]

T.Y., W.M.G., and J.F.L. led the project. 
W.M.G. proposed the follow-up optical observations and coordinated 
the spectroscopic and photometric data reduction and analysis, 
with significant inputs from T.Y., M.Y.S., Z.X.Z., S.W., Y.B., and L.L.Z.. 
Z.R.B. and H.T.Z. contributed to the early discovery of J1123. 
W.M.G., J.F.Wang, and J.F.Wu contributed to 
their expertise of the P200 proposals and observations. 
Y.B. and Z.X.Z. contributed to the P200 data reduction and analyses. 
T.Y., Q.Z.Y., and W.M.G. proposed the follow-up radio observations.
P.W. contributed to the expertise of FAST's data reduction and analyses.
Z.X.Z. and M.Y.S. solved the orbital solution using \texttt{PHOEBE}; 
S.W. and J.Z. validated the orbital solution independently.
T.Y., W.M.G., M.Y.S., J.F.L., and P.W. presented 
the physical interpretation of the data and wrote the manuscript.  
All authors reviewed and contributed to the manuscript.

\item[Competing interests] 
The authors declare no competing interests.

\item[Author information] 
Readers are welcome to comment on the online version of the paper. 
Correspondence and requests for materials should be addressed to 
W.M.G. (email: guwm@xmu.edu.cn), 
M.Y.S. (email: msun88@xmu.edu.cn),
or J.F.L. (email: jfliu@nao.cas.cn).

\end{addendum}

\clearpage
\begin{methods}\label{sec:method}

\subsection{Observations and their reduction}

I. Spectroscopy

Three low-resolution spectra (Supplementary Figure~1) 
were adopted from the first data-release of LAMOST low-resolution single epoch spectra\cite{Bai2021}.
Follow-up spectroscopy was obtained with DBSP mounted on the P200.
The first observation (eight exposures; 20 minutes for each) was taken during a
clear night on March 14, 2019, with an average seeing $\sim$ 2 arcsec.
The DBSP covered a wavelength range of 3800 -- 5400 \AA{} 
with a resolving power $R$ $=$ 3400 for blue arm and 
6000 -- 7600 \AA{} with $R$ $=$ 4900 for red arm 
(Supplementary Figure~2).  
We took HeNeAr and FeAr lamps for the wavelength calibration at the blue and red arms, respectively. 
Raw spectroscopic data was reduced by using the IRAF 
(Image Reduction and Analysis Facility).
Standard IRAF procedures of bias subtraction, flat correction, 
cosmic-rays removal, 1D-spectrum extraction, 
wavelength calibration, and flux calibration were conducted.
Wavelength calibration was performed by calibrating the HeNeAr and FeAr lamps
to the corresponding spectral atlas provided by 
the \href{http://iraf.noao.edu/specatlas/}{Spectral Atlas Center}.
After the reduction, a few cosmic rays (sharp spikes typically occupy only a few pixels) remained uncleaned, thus we removed them manually.
The second (remote) observation was taken on June 24, 2020 (two exposures),
with similar details described above. 
Supplementary Table~2 
is a complete observation log for spectroscopic observations. 

II. Photometry
 
J1123 was classified by the Catalina Real-time Transient Survey (CRTS)\cite{Drake2014} 
as a WUMa-type eclipsing binary,
but it was later noted to be an ellipsoidal variable\cite{Mu2022}.
We collect multi-band light curves from: the CRTS, 
the Zwicky Transient Facility (ZTF)\cite{Bellm2019}, 
the asteroid-hunting mission of the Wide-field Infrared Survey Explorer
(NEOWISE)\cite{Mainzer2011},
and the Transiting Exoplanet Survey Satellite (TESS)\cite{Ricker2015}.

J1123 was observed in sector 22 of TESS, 
with a monitoring time-span of roughly one month. 
The photometry of TESS was reduced by using the \texttt{Lightkurve} package\cite{Lightkurve2018}.
We used the \texttt{create\_threshold\_mask} method to create single aperture
and define background pixels.
We tried different TESS cutout sizes, aperture shapes, and background sizes 
to perform background subtraction and light curve extraction,
all reported well-consistent results. 
The simple background median subtraction method and also the 
\texttt{PLD} (pixel level de-correlation) method was used to extract the light curve. 

\subsection{The orbital period and ephemeris}

We derived the orbital ephemeris (Equation~(\ref{eq:ephem})) by using the data of CRTS\cite{Drake2014}.
The period was searched by using the Lomb-Scargle periodogram\cite{Lomb1976,Scargle1982},
with the generalized algorithm\cite{Zechmeister2009} and a fast implementation method\cite{Press1989}.
By following VanderPlas's prescription\cite{VanderPlas2018}, 
we set the frequency grid to be uniformly spaced for the periodogram 
with an oversampling factor of ten;
the lowest frequency is set to the reciprocal 
of the observation time-span $T_{\rm span}$,
while the highest frequency is chosen as 
500 times the average Nyquist frequency (i.e., Nyquist factor $=$ 500).
The Lomb-Scargle powers report a significant period component 
$P_{\rm peak}$ $=$ 0.13691772~days 
(Supplementary Figure~3; upper panel). 
The true orbital period is thus twice the $P_{\rm peak}$ as
the ellipsoidal light curve has two peaks and valleys,  
namely, $P_{\rm orb}$ $=$ 2$P_{\rm peak}$ $=$ 0.27383544(18)~days.
The uncertainty of $P_{\rm orb}$ 
(indicated by the last two digits inside the parenthesis)
was derived by finding periods for a collection of bootstrap-resampled light curves.
Periods for light curves of ZTF, NEOWISE, and TESS were searched
by using the same technique with similar details, 
except for that the Nyquist factor for TESS was set to be one
because of its 30-minutes observation cadence.
Lomb-Scargle powers reported consistent results for all surveys
(Supplementary Figure~3). 

To find the M star's superior conjunction, 
we folded the CRTS light curve with $P_{\rm orb}$ 
and a guessed reference HJD, denoted as HJD$_{\rm guess}$.
HJD$_{\rm guess}$ was taken to be the HJD of the point with the largest $V_{\rm mag}$ (smallest flux). 
Then the phase-folded light curve was fitted by a three-order Fourier series. 
The corresponding phase $\phi_{\rm min}$ at the best-fit's minimum ($V_{\rm mag}$ maximum) was found, 
which is supposed to coincide with the superior conjunction ($\phi$ $=$ 0) of the M dwarf.
The offset between the $\phi_{\rm min}$ and $\phi$ $=$ 0 helped us to
update (correct) the HJD$_{\rm guess}$ and hence to derive the reference HJD 
at superior conjunction as $T_{0}$ $=$ 2\,453\,734.909\,32 (31) HJD.

\subsection{The radial velocities fitting method}

To measure the RVs with respect to the Heliocentric rest-frame, 
we first use the \verb"baryvel" function provided by \texttt{PyAstronomy}'s \texttt{pyasl} package to obtain the Heliocentric corrections for the mid-time of exposures and the RVs. 
All spectra are then shifted by the corresponding Heliocentric  RV correction
(Supplementary Table~2; column 10). 
As mentioned, chromospheric emission lines 
are presented in all of J1123's spectra. 
We mask these emission lines
as well as strong telluric absorption lines 
in 6860\AA{} -- 6950\AA{} to measure the RV.

We use the \texttt{PyHammer} package\cite{Kesseli2017}
which implements the cross-correlation technique to measure the RVs
with a library of observational spectra collected from the SDSS BOSS survey.
The best-fitted template has a spectral type $=$ M1 and a metallicity $Z$ $=$ 0. 
Supplementary Figure~2 
shows the template fitting of
the last exposure taken by DBSP on Mar 14, 2019. 
The bootstrap re-sampling method is used
to measure the values and uncertainties for RVs.
For each spectrum, 3000 bootstrap re-samples are generated 
and fed to the \texttt{PyHammer}.  
The 50\%, 16\%, and 84\% percentiles of the result RVs are read 
as the normial value, the lower error bar, and the upper error bar, respectively.  
We also fit the first three Balmer lines H$\alpha$ -- H$\gamma$ 
to obtain their RV curves. 
The Gaussian function is used to fit the lines on the 
continuum-normalised spectrum.

Supplementary Table~2 
shows the measured RVs; 
Supplementary Figure~4 
shows comparisons for 
the RV curves of the first three Balmer lines 
(H$\alpha$, H$\beta$, and H$\gamma$) 
with the RV curve measured using only the absorption lines of the M star 
(i.e., all emission lines masked).
Note that all reported RVs are measured in the Heliocentric rest-frame.

\subsection{The spectral energy distribution fitting method}

We construct the broadband SED of J1123 using archival photometric measurements from 
GALEX\cite{Martin2005} (GALEX reported a detection of NUV $=$ 22.73 $\pm$ 0.52 mag but no detection of FUV), 
SDSS\cite{York2000} ($ugriz$ bands), 
{\it Gaia} EDR3\cite{Gaia2021} (BP, RP, and G bands), 
ZTF\cite{Bellm2019} ($gri$ bands), 
PanSTARRS\cite{Chambers2016} ($grizy$ bands), 
2MASS\cite{Cutri2003} (J, H, and Ks bands), 
and AllWISE\cite{Wright2010,Cutri2012} (W1--W4 bands).
The distance is $D = 318 \pm 4$ pc since the parallax 
$\bar \omega$ $=$ 3.147 $\pm$ 0.038~mas ({\it Gaia} EDR3\cite{Gaia2021}), 
and the extinction $E(B-V)$ $=$ 0.019 $\pm$ 0.001 
is referred to the PanSTARRS 3D dust extinction map\cite{Green2018}. 
Supplementary Table 3 
shows the collected archival photometric data.

We do not use GALEX NUV and SDSS u-band data when fitting SED,
since one does not have prior knowledge of the origin of these emissions. 
To measure the average temperature and the average radius of the M dwarf,
the mean flux is required in the SED fitting\cite{El-Badry2021}. 
Some of the surveys have sparse observations; for instance, 2MASS and SDSS 
both have observed J1123 only once. 
To account for the fact that these measurements are phase-dependent, 
we accommodate additional systematic flux uncertainties 
= $F_{\lambda} \times 5\% / \sqrt{N}$ for all bands, where 5\% is 
the standard deviation of the TESS light curve variations 
and $N$ is the number of observations for each band.
We also add additional 1\% uncertainties for all the ground-based surveys,
i.e., SDSS, PanSTARRS, ZTF, and 2MASS,
to account for the typical photometric zero-point uncertainties.

We used the Cardelli extinction curve\cite{Cardelli1989}
and the R$_\mathrm{V}$ $=$ 3.1 reddening law\cite{Fitzpatrick1999} 
to correct the reddening.   
A grid of synthetic photometry is constructed by integrating the transmission curves of the filters over 
the BT-settl\cite{Allard2012} synthetic model SED.
All synthetic SEDs and transmission curves were adopted from VOSA (virtual observatory SED analyzer)\cite{Bayo2008}.
The grid span of effective temperature is 3000 -- 5000~K, 
with a step size of 100~K;
and the span of surface gravity is 4.0 -- 5.5~dex, with a step size of 0.5~dex.
A Solar metallic abundance ($Z$ $=$ 0) is fixed by referring to 
the metallicity of the best-fitted \texttt{PyHammer} spectral template.   
A linear interpolation is used
(based on values at grid nodes) to approximate the synthetic photometry
at arbitrary points inside the pre-calculated grid. 

Prepared with the interpolated synthetic photometry grid, 
we use a MCMC sampler with the Metropolis-Hastings algorithm to fit the observed SED. 
We define the likelihood function $\mathcal{L}$ as
\begin{equation}
\ln \mathcal{L}(\mathcal{D}|\mathcal{M}(R_{1}, D, T_\mathrm{eff,~1}, \log{g})) =
-\frac{1}{2}
\sum_\mathrm{band}^{} \left(\frac{F_{\rm obs,band}-(\frac{R_{1}}{D})^{2}\times F_{\rm syn,band}(T_\mathrm{eff,~1}, \log{g})}{\sigma_{F}}\right)^{2} \ ,
\end{equation}
where $\mathcal{D}$ is the observed SED data set, $\mathcal{M}$ is the model SED given the parameters:
$R_{1}$ (stellar radius), $D$ (distance), $T_\mathrm{eff,~1}$ (effective temperature), and $\log{g}$ (surface gravity).
$F_{\rm obs,band}$ is the observed flux at a specific band, $(R_{1}/D)^{2}$ is the dilution factor,
$F_{\rm syn,band}(T_\mathrm{eff,~1}, \log{g})$ is the synthetic flux at a specific band, 
and $\sigma_{F}$ is the flux uncertainties.
Thus the MCMC sampler updates the prior distributions of our target parameters 
according to the \textit{Bayes} Theorem:
\begin{equation}
\mathcal{\pi}(\mathcal{M}(R_{1}, D, T_\mathrm{eff,~1}, \log{g})| \mathcal{D}) \propto \mathcal{L}(\mathcal{D}|\mathcal{M}(R_{1}, D, T_\mathrm{eff,~1}, \log{g})) \times \prod_{x \in X}^{} \mathcal{P}(x) \ ,
\end{equation}
where $\mathcal{\pi}$ is the posterior distributions of the parameters given the data;
$\mathcal{P}(x)$ is the prior distribution of a parameter $x$ 
in the target parameter set 
$X$ $=$ ($T_\mathrm{eff,~1}$, $R_{1}$, $D$, $\log{g}$).
The sampler uses the logarithm of the posterior, i.e., 
$\ln (\mathcal{\pi})$ in real implementation. 
The chain took 200\,000 steps with the first 100\,000 points dropped as the burn-in.
The prior distributions and posterior distributions of these parameters are listed in Supplementary Table~1. 
The marginal and joined posterior distributions 
are presented in Supplementary Figure~5. 

\subsection{The mass of the M dwarf}

Here we describe details of the isochrones fitting. 
The BT-Settl isochrones are again retrieved from the VOSA;
only isochrones with zero metallicity ($Z$ $=$ 0) are used.
These isochrones cover complete stellar ages, ranging from 
0.001 Gyr up to 12 Gyr. However, the mass grid is not fine enough,
e.g., only $M$ $=$ 0.45, 0.5, 0.57, 0.60, 0.62, 0.70, and 0.75 $M_{\odot}$ 
is provided for early-type M dwarfs or late-type K dwarfs. 
Thus a linear interpolation is implemented
to populate the mass grid in 0.45-0.75 $M_{\odot}$ mass range, 
with a finer step size of 0.01 $M_{\odot}$ bins.

By assuming the distribution is log-normal,
the likelihood for observing the M dwarf's effective temperature 
$T_\mathrm{eff,1}$ and bolometric luminosity $L_\mathrm{bol,1}$,
given a stellar mass $M_{*}$ can be calculated by:
\begin{equation}
\mathcal{L}(T_\mathrm{eff,~1}, L_\mathrm{bol,~1}|M_{*}) = 
\sum \frac{1}{2 \pi \sigma_{\log{T}} \sigma_{\log{L}}} 
\exp\left(-\frac{(\log T_\mathrm{eff,~1}-\log T_{{\rm eff},~M_{*}})^{2}}{2 \sigma_{\log{T}}^{2}} 
- \frac{(\log L_\mathrm{bol,~1} -\log L_{{\rm bol},~M_{*}})^{2}}{2 \sigma_{\log{L}}^{2}}\right) \ ,
\end{equation}
where the summation sign denotes the summation over all isochrones. 
$T_{{\rm eff},~M_{*}}$ and $L_{{\rm bol},~M_{*}}$ are 
the effective temperature and the bolometric luminosity
given the stellar mass $M_{*}$ in an isochrone,
and $\sigma_{T}$ and $\sigma_{L}$ are the uncertainties of 
$\log T_\mathrm{eff,~1}$ and $\log L_\mathrm{bol,~1}$, respectively. 
We adopted a Kroupa initial mass function\cite{Kroupa2001} 
as the prior of the stellar mass, denoted as IMF. 
Thus according to the \textit{Bayes} theorem, 
the probability distribution (posterior) 
of stellar mass $M_{*}$ given the observed
$T_\mathrm{eff,~1}$ and $L_\mathrm{bol,~1}$ is:
$\pi(M_{*}|T_\mathrm{eff,~1}, L_\mathrm{bol,~1})$ $\propto$
$\mathcal{L}(T_\mathrm{eff,~1}, L_\mathrm{bol,~1}|M_{*})$ $\times$ IMF.
The result posterior is obtained by fitting a Gaussian on the interpolated mass grid (Supplementary Figure~6; right panels). 

We obtain a stellar mass $M_{1}$ $=$ 0.60 $\pm$ 0.02 $M_{\odot}$ 
of the M dwarf by fitting the model 
BT-Settl isochrones. 
Note that this approach could have
uncertainty when dealing with a tidally locked star in close binaries.
According to previous studies\cite{Feiden2013}, 
stellar dynamo and magnetism\cite{Parker1955} can 
inflate the radii of low-mass stars ($<$0.8 M$_\odot$).
For tidally locked stars in close binaries, rapid rotation enhances 
the magnetic field\cite{Morgan2012} and induces more spots groups on the star's surface. 
Thus the star turns to be cooler than its 
single main-sequence (MS) siblings of the same type. 
The consequence is that the star is `bloated', the radius is inflated 
to balance the total output energy generated within the stellar interior.
Components in short-period ($\lesssim$1 day) binary systems can have radii being inflated by $\sim$4.8\% $\pm$ 1.0\%\cite{Kraus2011} 
with respect to single low-mass MS stars.
Thus we deflate 5\% of the SED derived radius.
Assuming that the bolometric luminosity is conserved 
during the tidal distortion process, this correction results in $\sim$2.6\% increase of the effective temperature 
(pink point in Supplementary Figure~6; left panel). 
The fitting of isochrone thus results in a slightly larger mass 
$M_{1}$ $=$ 0.61 $\pm$ 0.02 $M_{\odot}$. 

As an independent check,
we use an empirical mass-luminosity relation (MLR) for the MS M dwarfs 
(Equation (11) in Benedict et al. (2016)\cite{Benedict2016})
and the 2MASS Ks-band absolute Magnitude 
($=$ $K_\mathrm{mag}$ $+$ 5 $-$ 5$\log D$ $-$ $A_\mathrm{K}$ $=$ 5.18 $\pm$ 0.04, 
where $K_\mathrm{mag}$ $=$ 12.699 $\pm$ 0.023 and 
$A_\mathrm{K}$ $=$ 0.0068 $\pm$ 0.0004) to estimate the mass of the visible M dwarf. 
We obtain $M_{1} = 0.62 \pm 0.02 M_{\odot}$, 
which is similar to the result from the model isochrone fitting.  
We adopt the isochrone mass 
$M_{1}$ $=$ 0.61 $\pm$ 0.02 $M_{\odot}$, 
as the mass of the M dwarf.

\subsection{The orbital solution by the \texttt{PHOEBE} software}

To obtain the orbital inclination $i$, 
the mass ratio $q\, (\equiv M_{2}/M_{1})$, 
and thus the compact object's mass $M_{2}$,
we fit simultaneously the RV curve and the TESS light curve using 
\texttt{PHOEBE 2.3}\cite{Prsa2005,Prsa2016,Conroy2020}.
The orbital period ($P_{\rm orb}$ $=$ 0.27383544 days) is fixed. 
For the compact object (secondary), we set \texttt{distortion\_method $=$ none}
such that it is treated like an object without flux contributions 
and eclipse effects. 
For the M dwarf (primary), the effective temperature 
($T_\mathrm{eff,~1}$ $=$ 3769 K) is fixed.
We obtain the limb-darkening($u$) coefficient
and the gravity-darkening($\tau$) coefficient 
in TESS band by using Claret 2017\cite{Claret2017}'s 
limb-darkening and gravity-darkening tables.  
The results are: $\tau$ $=$ 0.32 for gravity-darkening,
and ($e$, $f$) $=$ (0.684, 0.389) for a logarithmic limb-darkening 
law (refer to Equation (4) in Claret 2017\cite{Claret2017}).
The \texttt{PHOENIX} atmospheric model is used.

The first model (hereafter model A) we considered is the canonical 
ellipsoidal modulations\cite{Morris1985,Morris1993,Gomel2021,Rowan2021} with $M_1=0.61\ M_{\odot}$ and $R_1=0.63\ R_{\odot}$. 
Given $i$ and $M_2$, the logarithmic likelihood of the observed TESS light 
curve is 
\begin{equation}
\label{eq:phoebe_like_lc}
    \ln \mathcal{L}(\mathrm{TESS\ LC}|i, M_2, \mathrm{model\ A})=
    -\frac{1}{2}\sum \left[\frac{(f_{\mathrm{m}} - f_{\mathrm{obs}})^2}{\sigma_{\mathrm{tot}}^2} 
    +
    \ln{(2\pi\sigma_{\mathrm{tot}}^2)}\right] \\,
\end{equation}
where $f_{\mathrm{m}}$ and $f_{\mathrm{obs}}$ represent the model and 
observed \texttt{TESS} fluxes, respectively; $\sigma_{\mathrm{tot}}^{2} 
=\sigma_{\mathrm{obs}}^{2} + \sigma_{\mathrm{sys}}^{2}$ and the latter two 
represent the uncertainties of the observed \texttt{TESS} fluxes 
and the systematic uncertainties (which accounts for additional 
factors, e.g., starspot, that were not included in the model). The 
ellipsoidal modulation fluxes are calculated via \texttt{PHOEBE}. 
Similarly, we can calculate the logarithmic likelihood of the observed 
RV variations for every $i$ and $M_2$, 
i.e., 
\begin{equation}
\label{eq:phoebe_like_rv}
    \ln \mathcal{L}(\mathrm{RV_{obs}}|i, M_2, \mathrm{model\ A})=
    -\frac{1}{2}\sum \left[\frac{(v_{\mathrm{m}} - v_{\mathrm{obs}})^2}{\sigma_v^2} 
    +
    \ln{(2\pi\sigma_v^2)}\right] \\,
\end{equation}
where $v_{\mathrm{m}}$, $v_{\mathrm{obs}}$ and $\sigma_v$ represent 
the model RV, the observed RV and its uncertainty. 
The prior distributions for the model parameters are assumed to be 
uniform. The Markov Chain Monte Carlo Ensemble sampler code 
\texttt{emcee}\cite{Foreman-Mackey2013} is used to sample the posterior 
probability density, i.e., the product of the likelihood (whose logarithmic 
value is the summation of Equations~\ref{eq:phoebe_like_lc} and 
\ref{eq:phoebe_like_rv}) and the uniform prior probability. 
We implement 50 parallel chains, each with 
10\,000 steps. 
The results are $i=73^{+1.8}_{-1.5}$ degree and 
$M_2 = 1.24^{+0.03}_{-0.03}\ M_{\odot}$. The fitting result is presented 
in Supplementary Figure~7. 
It is clear that the residuals between the 
model and the observed fluxes show periodic variations. The Akaike 
Information Criterion (AIC) and Bayesian Information Criterion (BIC) for 
this fit are $-348$ and $-337$, respectively. 

Motivated by the periodic variations in the residuals, we consider 
the following two models: a model with ellipsoidal modulations and a 
hotspot (hereafter model B); a model with ellipsoidal modulations 
and a coldspot (hereafter model C). We find that models B and C eliminate 
the periodic residuals (see Supplementary Figure~7).  
The AICs and BICs for the two models are smaller 
than those of model A by a factor of $60$. Statistically speaking, models 
B and C are preferred over model A. To fully propagate the uncertainties 
of $M_1$ and $R_1$ to our estimation of $i$ and $M_2$, we now also let 
$M_1$ and $R_1$ in models B and C to vary; we set Gaussian priors for $M_1$ 
and $R_1$ according to the isochrone mass and the SED radius, i.e., 
$\mathcal{N}(0.61 M_{\odot},0.02 M_{\odot})$ for $M_{1}$ 
and  $\mathcal{N}(0.63 R_{\odot},0.01 R_{\odot})$ for $R_{1}$. 
For model B, the results are $i=71_{-7.2}^{+10}$ degree and 
$M_2=1.26_{-0.09}^{+0.13}\ M_{\odot}$; for model C, the results are 
$i=64_{-4.9}^{+6.3}$ degree and $M_2=1.4_{-0.11}^{+0.12}\ M_{\odot}$. 
Models B and C cannot be distinguished since their AICs and BICs 
are similar. We stress that the three inferred $M_2$ are statistically 
consistent (within $1.5$-sigma uncertainties).

\subsection{Could a distant third body exist in J1123?}

Triple systems are not rare in the Galaxy.
Could a distant third body (i.e., the tertiary) exist in J1123? 
To address the issue, we use two techniques that are popular for 
characterizing spectroscopic binaries or 
triples\cite{Merle2017,Cunningham2019,Traven2020,Kounkel2021,Li2021}: 
the cross-correlation function (CCF)\cite{Tonry1979} and 
the broadening function (BF).\cite{Rucinski1992,Rucinski1999} 
We create a theoretical template having the same stellar parameters 
as J1123, by interpolating the theoretical BT-settl models\cite{Allard2012} 
for low-mass stars.
Using the template, the CCFs and the BFs for P200 spectra are calculated
in the velocity range 
$(-500 \mathrm{\, km\, s^{-1}}$, $500 \mathrm{\, km\, s^{-1}})$
and for three wavelength ranges: 
4360 -- 4800~\AA{}, 5190 -- 5360~\AA{}, and 6050 -- 6280~\AA{} (as they 
are free of emission lines, telluric absorption, and provide fine matches 
to the theoretical template). The final result is obtained by averaging 
the results of these three wavelength ranges 
(Supplementary Figure~8). 

For comparison, we create mock data using the same BT-Settl template, 
to simulate the spectrum for a triple system scenario with a distant third star plus an inner binary (M star + compact object). 
Only the wavelength range 6000 -- 7000~\AA{} is used for simplicity. 
For the M star, we artificially broaden the spectrum with a rotational broadening kernel\cite{Gray2005} corresponding to 
$v\sin{i} = \omega R_{1} \sin{i} = (2 \pi / P_{\rm orb}) R_{1} \sin{i} \simeq 111 \mathrm{\, km\, s^{-1}}$. The spectrum is then shifted to the corresponding RV at a specific phase. 
For the third star, we assume that it is a slow rotating one; 
a rotational broadening kernel with $v\sin{i} = 10 \mathrm{\, km\, s^{-1}}$ is applied. 
We also assume that the third star's RV is sufficiently small, i.e.,  
$V_{\rm R} \simeq \gamma = -8 \mathrm{\, km\, s^{-1}}$ 
in the case of a very wide outer orbital separation or the case of a nearly face-on outer orbit. The luminosity of the third star is assumed to be $0.5$, $0.2$, or $0.1$ 
times that of the M dwarf.
The two mock spectra are superimposed, degraded to a resolution $R = 4900$ similar to our P200 spectra, 
and Gaussian noise is added according to the signal-to-noise ratio of the corresponding real spectrum. The CCFs and the BFs are then calculated by using the mock spectra and the template from which these mock spectra are generated.

A prominent secondary peak is expected in the CCFs and BFs of the mock data if the 
distant third object is contributing a non-negligible ($\gtrsim 10\%$) fraction of flux 
(lower sub-panels of Supplementary Figure~8). 
On the contrary, both CCFs and BFs for J1123 observations show single-peaked shapes, 
in well agreement with a single-lined spectroscopic binary. 
The comparison between the CCFs and BFs of real observations and those of our mock data 
suggests that the possible tertiary should be one order of magnitude fainter than the 
visible M dwarf. 

In addition, high precision \textit{Gaia} astrometry measured 
an astrometric excess noise $\epsilon = 0 \pm 0$~mas 
and a renormalized unit weight error\cite{Lindegren2021} \texttt{ruwe} $=$ 1.008. 
Instead, wide binaries or triple systems in the Solar neighbourhood 
tend to have a significantly larger excess noise\cite{Gandhi2022} 
or a \texttt{ruwe} significantly deviates from one\cite{Belokurov2020}.
Hence, the two \textit{Gaia} astrometric parameters suggest that 
J1123's astrometric solution is inconsistent with the triple 
system scenario. 

\subsection{The H$\alpha$ luminosity, UV excess, and the chromospheric activity}
To obtain the H$\alpha$ luminosity,
we scale the target spectra to match the model SED in the first place. 
The model SED at a wavelength range 
6564.61 $\pm$ 30 \AA{} around the H$\alpha$ is subtracted
and the total H$\alpha$ flux  $F_\mathrm{H\alpha}$ can then be summed up.
The H$\alpha$ luminosity $L_\mathrm{H\alpha}$ $=$ 
$F_\mathrm{H\alpha} \times 4 \pi R_{1}^{2} \times (R_{1}/D)^{-2}$,
i.e., $\simeq$ $2\times$$10^{28}$ erg~s$^{-1}$,
or equivalently, $\simeq$ 5.3$\times$$10^{-6}$$L_{\odot}$ (average over all 13 spectra).

J1123 was recognized to be an UV emitting star by Bai et al. (2018)\cite{Bai2018}.
GALEX's All-sky Imaging Survey (AIS) has observed J1123 on Feb 14, 2007.
Both NUV and FUV filters took 136 seconds' exposure simultaneously, 
but only NUV has reported detection of NUV $=$ 22.73 $\pm$ 0.52 mag (AB magnitude system).
GALEX's observation was taken at the orbital phase $\phi$ $=$ 0.248,
near the quadrature-phase when two components were side by side. 
Thus the compact object was not obscured by the M dwarf 
but being fully exposed to the observer.

The extinction corrected GALEX NUV flux 
$F_\mathrm{NUV}$ $=$
(1.9 $\pm$ 0.9) $\times$ 10$^{-17}$~erg~s$^{-1}$~cm$^{-2}$~\AA{}$^{-1}$,
thus the NUV luminosity $L_\mathrm{NUV}$ $\approx$
$F_\mathrm{NUV}$$\times$$\mathrm{W_{eff}}$$\times$$(R_{1}/D)^{-2}$$\times$4$\pi$$R_{1}^{2}$ $=$ (1.8$\pm$0.8)$\times$10$^{29}$erg s$^{-1}$,
where $\mathrm{W_{eff}}$ $=$ 768.31\AA{} is the effective width of NUV.
We calculate the excess of UV luminosity by subtracting 
the base UV emission from the photosphere. 
The latter is obtained by integrating the model SED in the NUV band,
which yields $L_\mathrm{base}$ $\approx$ 1.0$\times$10$^{28}$ erg s$^{-1}$.
Thus, the UV excess can be calculated as $R_\mathrm{NUV}$ 
$=$ ($L_\mathrm{NUV} - L_\mathrm{base}$) $/$ $L_\mathrm{bol}$ 
$=$ 0.00063 $\pm$ 0.00032 and $\log{R_\mathrm{NUV}}$ $=$ $-$3.20$\pm$0.22.
The empirical $L_\mathrm{H\alpha}$ - $L_\mathrm{NUV}$ relation\cite{Jones2016}
predicts that 
$\log$($L_\mathrm{NUV}/L_\mathrm{bol}$) $=$ 0.67$\times$ 
$\log$($L_\mathrm{H\alpha}/L_\mathrm{bol}$) $-$ 0.85 $\approx$ $-$3.5
(or equivalently, 8.8 $\times$ 10$^{-18}$ erg s$^{-1}$ cm$^{-2}$ \AA\ $^{-1}$), 
provided that 
$\log$(${\overline L_\mathrm{H\alpha}}/L_\mathrm{bol}$) $\approx$ $-$4.

The UV spectra of AD Leo and AU Mic, two single M dwarfs, 
were retrieved from the Mikulski Archive for Space Telescopes 
(MAST\footnote{\url{https://archive.stsci.edu/index.html}}).
Both sources were observed by the IUE (International Ultraviolet Explorer)\cite{Boggess1978a,Boggess1978b} multiples times.
For each source, we collect all the low dispersion spectra, 
take the average spectrum, and convolve the average by an 1 \AA\ kernel to increase the signal-to-noise ratio. 
We scale the UV spectra according to (a similar method 
described by Rugheimer et al. (2015)\cite{Rugheimer2015} ):
\begin{equation}
F_\mathrm{UV~J1123} =
F_\mathrm{UV~AD/AU} \times \left(\frac{T_\mathrm{eff~J1123}}{T_\mathrm{eff~AD/AU}}\right)^{4} 
\times \frac{\mathrm{log}(L_\mathrm{H\alpha~J1123}/L_\mathrm{bol~J1123})}
{\mathrm{log}(L_\mathrm{H\alpha~AD/AU}/L_\mathrm{bol~AD/AU})} 
\times \left(\frac{R_\mathrm{J1123}}{R_\mathrm{AD/AU}} \times \frac{D_\mathrm{AD/AU}}{D_\mathrm{J1123}}\right)^{2} \ ,  
\end{equation}
where the subscript AD/AU stands for AD Leo or AU Mic.
The adopted stellar parameters for two sources and references are listed in the Supplementary Table~4. 
The synthetic NUV flux from the chromosphere is then calculated 
by integrating the UV spectrum with GALEX NUV transmission curve.
The mean NUV flux of two sources is adopted as the final estimated synthetic photometry.

\subsection{FAST observation of J1123 and data reduction}
The FAST observation of J1123 was conducted on September 7, 2021. The total integration time was 3000 seconds (the first two minutes and the last two minutes was for calibration signal injection time).
The centre frequency was 1.25 GHz, spanning from 1.05 GHz to 1.45 GHz, including a 20-MHz band edge on each side. The average system temperature was 25 K. The recorded FAST data stream for pulsar observations is a time series of total power per frequency channel, stored in PSRFITS format (Hotan et al., 2004)\cite{Hotan2004} from a ROACH-2\footnote{\url{https://casper.ssl.berkeley.edu/wiki/ROACH-2\_Revision\_2}} based backend, which produces 8-bit sampled data over 4k frequency channels at 49 $\mu$s cadence. We searched for radio pulsations with either a dispersion signature or instrumental saturation in all FAST data collected during the observational campaign. Three types of data processing were performed: I) dedicated (half-blind) periodic pulse search, single pulse search and baseline saturation search.

I. Dedicated (half-blind) search:

Based on the Galactic electron density model NE2001\cite{Cordes2002} and YMW16\cite{Yao2017}, 
we estimate the distance $D$ $=$ 318 $\pm$ 4 pc corresponding with a DM of $\sim$ 3.2 pc cm$^{-3}$, 
and the line of sight maximal Galactic DM (max) $=$ 12 pc cm$^{-3}$. 
Due to model dependence and for the sake of robustness, 
we set the range of dispersion search as approximately zero to four times the estimated DM (max), 
namely, DM $=$ 0 -- 50 pc cm$^{-3}$, which should cover all uncertainties. 
We used the output of PRESTO's \texttt{DDPlan.py} package 
(Ransom et al., 2001\cite{Ransom2001})
to establish the de-dispersion strategy. 
The step size between subsequent trial DMs 
($\Delta$DM = 0.1 pc cm$^{-3}$) was chosen 
such that over the entire band $t$($\Delta$DM) $=$ $t_{\rm channel}$. 
This ensures that the maximum extra smearing caused by 
any trial DM deviating from the source DM by $\Delta$DM 
is less than the intra-channel smearing.
For each of the trial DMs, we searched for a periodical signal 
and the first and second order 
(jerk search; Andersen \& Ransom, 2018 \cite{Andersen2018}) 
acceleration in the power spectrum based on the PRESTO pipeline (Wang et al. 2021)\cite{Wang2021}. 
We checked all the pulsar candidates of signal-to-noise-ratio (SNR) $>$ 5 
one by one and identified them as narrow-band radio frequency interferences (RFIs). 

II. Single pulse search:

We used the above dedicated search scheme to de-disperse the data. Then we used 14 box-car width match filter grids distributed in logarithmic space from 0.1 ms to 30 ms. A zero-DM matched filter was applied to mitigate RFI in the blind search. All the possible candidate plots generated were then visually inspected. Most of the candidates were RFIs, and no pulsed radio emission with dispersive signature was detected with a SNR $>$ 5. 

III. Saturation search:

We understand that if the radio flux is as high as kilojansky–megajansky, FAST would be saturated. We therefore also searched for saturation signals in the data. We looked for the epoch in which 50$\%$ of channels satisfy one of the following conditions: 1) the channel is saturated (255 value in 8-bit channels), 2) the channel is zero-valued, 3) the RMS of the bandpass is less than 2. We did not detect any saturation lasting $>$ 0.5 s, hence excluded any saturation associated during the observational campaign.

IV. Notes on the acceleration searches:

Since the orbital period (6.6 hr) is longer than about 8 times the observation duration (3000 sec), we carried 
out acceleration searches for the binary pulsar candidate using line acceleration ($Z_{\rm max}=$300) and jerk 
($W_{\rm max}=$100) terms in the Fourier domain (PRESTO). 
This corresponds to a maximum drift of physical linear acceleration 
$Z_{\rm max} \cdot c \cdot P_{\rm spin}/N_{\rm harm}/T_{\rm obs}^{2}$,
where c is the speed of light, $P_{\rm spin}$ is the spin period of the pulsar, and $T_{\rm obs}$ = 3000 s is the effective integration time. 
We note that all acceleration and jerk effects affect higher harmonics of a pulsar signal more than they do the fundamental ones. 
For a non-recycled NS, assuming the typical $P_{\rm spin} =$ 20 and 200~ms pulsar detected with up to eight harmonics ($N_{\rm harm} = 8$, incoherently summarizing possible harmonics in fundamentals to increase the SNR),
the maximum linear accelerations are $\sim$25 and $\sim$250~m~s$^{-2}$, respectively. 
That is, the range of linear acceleration are ($-$25, 25) and ($-$250, 250)~m~s$^{-2}$.
For the ‘jerk’ search, the constant jerk corresponds to a 
linearly varying acceleration: 
$W_{\rm max} \cdot c \cdot P_{\rm spin}/N_{\rm harm}/T_{\rm obs}^{3}$, hence the maximum jerk acceleration 
can be separately approximated as $\sim$0.003 and $\sim$0.03 m s$^{-3}$, 
i.e., the range of jerk acceleration are 
($-$0.003, 0.003) and ($-$0.03, 0.03)~m~s$^{-3}$, 
assuming the same $P_{\rm spin} =$ 20 and 200~ms.

We note that the acceleration and the jerk searching effects highly depend on the orbital phase of the NS. 
For J1123, our FAST observation corresponds to
the M star's phases in between $\phi=$ 0.24 -- 0.37, 
or equivalently, the NS's phases in between $\phi=$ 0.74 -- 0.87.
Utilizing the well-measured RV curve of the M star, the inferred RV curve for the NS can be written as: $V_{\rm R, NS} = (K_{1} / q) \cdot \sin((2 \pi / P_{\rm orb}) t) + \gamma$, where $q = 2.04 \pm 0.04$ is the dynamically constrained mass ratio.
We estimated the variation of the linear acceleration and the jerk of the NS using the first and the second derivative of the NS's RV curve:
$V'_{\rm R, NS} = (K_{1} / q) \cdot (2 \pi / P_{\rm orb}) \cdot \cos((2 \pi / P_{\rm orb}) t)$
and $V''_{\rm R, NS} = - (K_{1} / q) \cdot (2 \pi / P_{\rm orb})^{2} \cdot \sin((2 \pi / P_{\rm orb}) t)$, respectively.
Thus the range of the linear acceleration and the jerk are 
(2.1$\pm$0.1, $-$22.9$\pm$0.4)~m~s$^{-2}$ and 
($-$0.0089$\pm$0.0002, $-$0.0065$\pm$0.0002)~m~s$^{-3}$, respectively. 
Based on our estimation above, the parameter space for pulsar search 
is sufficient for the radio pulse search of this binary system.

\subsection{The proper motion and the Galactic position of J1123}

The proper motion of J1123 is\cite{Gaia2021} 
$\mu_{\alpha} = -20.604 \pm 0.034\, \mathrm{mas\, yr}^{-1}$ 
in the R.A. direction and 
$\mu_{\delta} = -24.276 \pm 0.036\, \mathrm{mas\, yr}^{-1}$
in the Dec. direction,
thus the total proper motion is
$\mathrm{pm} = 31.841 \pm 0.035\, \mathrm{mas\, yr}^{-1}$.
Located at a distance $D = 318 \pm 4$ pc, 
the transverse velocity on the projected plane of the sky is
$V_{T} \approx \mathrm{pm} \times D = 48.0 \pm 0.6\, \mathrm{km\, s}^{-1}$
and the space velocity is
$V = \sqrt{V_{T}^{2} +  V_{R}^{2}} = 48.7 \pm 0.7\, \mathrm{km\, s}^{-1}$.

The Galactic longitude and latitude of J1123 is 
$l =$ 171.85948 degree and $b =$ 67.58776 degree, respectively,
placing it on a height $z \approx z_{\odot} + D \sin{b}$ $\simeq$ 319 pc above the Galactic plane
(where $z_{\odot}$ $=$ 25 pc is the height of our Sun above the Galactic plane\cite{Juric2008}). 
By using the \texttt{pyasl} package in \texttt{PyAstronomy},
we calculate J1123's Galactic space velocity relative to the local standard rest  
as $(U, V, W) = (-1.4~{\rm km~s^{-1}}, -33.5~{\rm km~s^{-1}}, -3.8~{\rm km~s^{-1}})$.
We use the method described by Bensby et al. 2003\cite{Bensby2003}
to calculate the relative probability for the thick-disc-to-thin-disc membership, qualified as TD/D. We find that TD/D = 0.028, 
that is, J1123 most likely belongs to the thin disc.

\subsection{Characterizing the possible massive, cold WD}

As mentioned in the paper, the possibility that a massive cold WD resides 
in J1123 cannot be fully ruled out. 
Here, we constrain the temperature of of the possible WD.
We adopt theoretical WD atmospheric models\cite{Koester2010}
plus the best-fit M dwarf's SED to construct composite SEDs
(Supplementary Figure~9). 
Four DA type WD models with a same surface gravity $\log{g}$ $=$ 9.0 
but different effective temperatures 
($T_\mathrm{eff, WD}$ $=$ 50\,000 K, 20\,000 K, 10\,000 K, and 8000 K) are adopted.
The WDs' radius $R_{\rm WD}$ $\approx$ 0.0057 $R_{\odot}$ corresponds to the 
mass $M_\mathrm{WD} = $1.2$ M_{\odot}$.
The WD SEDs are scaled to J1123's distance ($D = 318$ pc) by multiplying the dilution factor $(R_{\rm WD}/D)^{2}$.

Supplementary Figure~9 
compares the composite SED (purple curve) and its synthetic flux 
(hollow purple diamonds) with the observed P200 spectrum and photometric fluxes. 
WDs with $T_\mathrm{eff, WD}$ $=$ 50\,000 K or 20\,000 K are ruled out
since models predict higher fluxes
than the observations at the GALEX NUV band or SDSS u-band. 
A WD with $T_\mathrm{eff, WD}$ $\approx$ 10000K can account for the NUV flux. 
In the realistic case, since the SED contains a substantial amount of 
chromospheric UV emission, the possible WD is even colder, that is, $T_\mathrm{eff, WD} <$10\,000K.

A recently discovered detached post-common-envelope binary (PCEB)
SDSS J1140+1542\cite{Parsons2021} (J1140 for short)
contains a similar massive cold WD with 
$M_\mathrm{WD} > 1.22M_{\odot}$ and $T_\mathrm{eff~WD} = 8900 \pm 900$K.
The companion of J1140 has a spectral type M2.5,
a slightly cooler M dwarf with $T_\mathrm{eff}$$\sim$3400 K, 
and only half of J1123's luminosity 
(i.e., $L_\mathrm{bol~J1140}$$\sim$0.035$L_{\odot}$) as reported by \textit{Gaia}.
Due to its relatively faint companion, 
the WD nature of J1140 was able to be revealed by the spectral decomposition method.
In fact, a handful of magnetic WDs\cite{Liebert2005(b),Schreiber2021} 
in detached PCEBs were found to be systematically much cooler 
($T_\mathrm{eff~WD}\lesssim10000$K), more massive, 
and close to Roche lobe filling than non-magnetic ones\cite{Parsons2021}.
But again, these magnetic systems are lower in companion masses
as their spectral types are typically later than M3.0.  
Therefore if the compact object in J1123 is a WD, 
the WD must be cold and massive, 
which can be hardly found by conventional (non-time-domain) WD surveys.


\subsection{Investigating the irradiation effects}

I. NS + irradiating M star case:

We simulate the TESS light curve for the case of a NS + M star binary
with irradiation effects using \texttt{PHOEBE} 
(the left panels of Supplementary Figure~10). 
We use 100 sets of orbital parameters randomly drew from 
the distributions of the \texttt{PHOEBE}’s orbital solution to 
simulate the light curves.

We consider three irradiation luminosities, 
$L_{\mathrm{irr}}=$ $4.2\times 10^{32}\ \mathrm{erg\ s^{-1}}$, 
$6.4\times 10^{31}\ \mathrm{erg\ s^{-1}}$, 
and $1.6\times 10^{30}\ \mathrm{erg\ s^{-1}}$. 
It is evident that for $L_{\mathrm{irr}} \gtrsim 6.4 \times 10^{31}\ \mathrm{erg\ s^{-1}}$, 
the resulting model light curve is inconsistent with the TESS observations (Supplementary Figure~10; left upper and middle panels). 
The nose of the M star (the side facing the inner Lagrange point) 
is heated up by the incident energy,
partitioning the M star into a noticeable ``day-side'' (near $\phi = 0.0$) 
and a ``night-side'' (near $\phi = 0.5$ ). Hence, we conclude that $L_{\mathrm{irr}}\lesssim 6.4\times 10^{31}\ \mathrm{erg\ s^{-1}}$. 

The energy source of the irradiation is probably powered by the pulsar wind of the NS. About $20\%$\cite{Breton2013} of the pulsar wind power is reprocessed and heats up the M dwarf. 
Hence, according to the analyses above,
the wind power of J1123 should be 
$L_{\mathrm{wind}}\lesssim 3.2\times 10^{32}\ \mathrm{erg\ s^{-1}}$. 
The possible pulsar wind in J1123 should be driven by 
the spin-down energy of the NS 
(rather than the accretion power since J1123 is a non-accreting system). Therefore, the spin-down power $L_{\mathrm{sd}}$ should also be less than $3.2\times 10^{32}\ \mathrm{erg\ s^{-1}}$. 
The spin-down rate of a NS is determined by
$\dot{P}_{\rm spin} = (L_{\rm sd} P_{\rm spin}^{3})/(4 \pi^{2} I_{\rm NS})$,
where 
$I_{\rm NS} = (2/5) \cdot M_{\rm NS} R_{\rm NS}^{2} \simeq 1.4 \times 10^{45} ~\mathrm{g~cm^{2}}$ is the moment of inertia, 
assuming that the NS is an uniform sphere with a mass
$M_{\rm NS} = 1.24 M_{\rm \odot}$ and a radius $R_{\rm NS} = 12~\mathrm{km}$. 
We further constrain the spin-down rate of J1123 to be
$\dot{P}_{\rm spin} \lesssim 4.6 \times 10^{-20} - 4.6 \times 10^{-17}
~\mathrm{s/s}$ assuming the NS is a non-recycled one 
with $P_{\rm spin} = 20 - 200~\mathrm{ms}$.
The constraint indicates a NS age $\gtrsim 10^{8} - 10^{10}$ yr.

II. WD + M star case: 

We simulate the ZTF g-band light curve for the case of a WD + M star binary 
with irradiation effects using \texttt{PHOEBE} 
(the right panels of Supplementary Figure~10). 
We use 100 sets of orbital parameters randomly drew from 
the distributions of the \texttt{PHOEBE}’s orbital solution to 
simulate the light curves.

We examine WDs with 
$T_{\rm eff,~WD} = 20000\,\mathrm{K}$, 
$10000\mathrm{K}$, and $8000\mathrm{K}$.
The radius of the WD is set to be $0.005 R_{\odot}$. 
For a hot WD with $T_{\rm eff,~WD} = 20000\,\mathrm{K}$, 
when the orbital inclination angle is large (nearly edge-on case), a prominent eclipse is expected at the M star’s inferior conjunction
phase $\phi = 0.5$ (Supplementary Figure~10; right upper panel);  
otherwise, there are no eclipses in the simulated light curves. 
For a WD with $T_{\rm eff,~WD} \lesssim  10000\,\mathrm{K}$,
possible eclipse becomes hard to be detected given the photometric precision 
of the ZTF light curve 
(Supplementary Figure~10; right middle and lower panels). 
In summary, our two simulations suggest that the invisible object is either a NS or a cold massive WD.

\end{methods}

\clearpage



\renewcommand{\tablename}{Supplementary Table}
\renewcommand\thetable{\arabic{table}}   
\setcounter{table}{0}

\begin{table}
\renewcommand{\arraystretch}{1.05}	
\centering
\scriptsize

\begin{tabular}{@{}lllll@{}}
\hline
\textbf{Parameter} & \textbf{Units} & \textbf{Prior} & \textbf{Posterior / Value} & \textbf{Notes} \\
\hline
\multicolumn{4}{l}{\bf Astrometric information$^{[18]}$}\\
\hline
RA & deg (J2000) &  -- & 11: 23: 06.93 & Right Ascension\\
Dec & deg (J2000) &  -- & +40: 07: 36.75 & Declination \\
parallax & mas & -- & 3.147 $\pm$ 0.038 & Trigonometric parallax measured by {\it Gaia} EDR3\\
$D$ ({\it Gaia}) & pc & -- & 318 $\pm$ 4 & Distance derived from {\it Gaia} EDR3 parallax\\
$\mu_{\alpha}$ & mas yr$^{-1}$ 
& -- & -20.604$\pm$0.034 
& Proper motion in RA measured by {\it Gaia} EDR3 \\
$\mu_{\delta}$ & mas yr$^{-1}$ 
& -- & -24.276$\pm$0.036 
& Proper motion in Dec measured by {\it Gaia} EDR3 \\
$l$ & degree 
& -- & 171.85948 
& Galactic longitude by {\it Gaia} EDR3 \\
$b$ & degree 
& -- & 67.58776 
& Galactic latitude by {\it Gaia} EDR3 \\
\texttt{ruwe} & -- & -- & 1.008 & Renormalised unit weight error by {\it Gaia} EDR3\\
\texttt{excess noise} & mas & -- & 0 $\pm$ 0 & Astrometric excess noise by {\it Gaia} EDR3\\
\hline
\multicolumn{4}{l}{\bf Photometric and spectroscopic information}\\
\hline
$T_{0}$($\phi$ $=$ 0) & HJD & -- & 2\,453\,734.909\,32(31) & The superior conjunction HJD of the M dwarf \\
$K_1$ & $\mathrm{km\,s^{-1}}$ & $\mathcal{U}(-500, 500)$ & 257 $\pm$ 2 & Semi-amplitude of the M dwarf's RV curve\\
$\gamma$ & $\mathrm{km\,s^{-1}}$ & $\mathcal{U}(-100, 100)$ & $-$8 $\pm$ 2 & Systemic RV of J1123 \\
$E(B-V)$ & mag & -- & 0.019 $\pm$ 0.001 & Interstellar reddening by PanSTARRS 3D dust-map$^{[49]}$\\
\hline
\multicolumn{4}{l}{\bf Stellar parameters of the visible M dwarf}\\
\hline
$R_1$ (SED) & \rsun & $\mathcal{U}(0.1, 1.0)$ & 0.63 $\pm$ 0.01 & Volume-averaged radius of the M dwarf by fitting SED\\
$D$ (SED) & pc & $\mathcal{N}(318,4)$$^{[18]}$ & 318 $\pm$ 4 & Distance derived from fitting the SED\\
$T$$_\mathrm{eff,1}$ (SED) & K &  $\mathcal{U}(3000, 5000)$  & 3769$_{-9}^{+8}$ & Mean stellar effective temperature by fitting SED\\
$L$$_\mathrm{bol,1}$ (SED) & $L_{\odot}$ &  --  & 0.0694 & Bolometric luminosity by integrating the model SED\\
$\log{g}_1$ (SED) & dex & $\mathcal{N}(4.67, 0.10)^{(*)}$ & 4.88 $\pm$ 0.08 & Surface gravity by fitting SED\\
$L$$_\mathrm{bol,1}$ & $L_{\odot}$ &  --  & 0.0710 $\pm$ 0.0021 & Bolometric luminosity by 
$L_\mathrm{bol}$ $=$ 4 $\pi R^{2} \sigma T_\mathrm{eff}^{4}$\\
$M_1$ (ISO) & \msun & IMF$^{[55]}$ & 0.61 $\pm$ 0.02 & Mass of the M dwarf by fitting model isochrone\\
$\log{g}_1$ (ISO) & dex & -- & 4.63 $\pm$ 0.02 & Surface gravity by fitting stellar evolution model\\
$M_1$ (MLR) & \msun & -- & 0.62 $\pm$ 0.02 & Mass of the M dwarf by using the IR MLR$^{[60]}$\\
\hline
\multicolumn{4}{l}{\bf Orbital solutions by 
\texttt{PHOEBE}$^{[61-63]}$ 
(Model A)}\\
\hline
$q$ & $\equiv M_{2} / M_{1}$ & -- & $2.04\pm0.04$ & Mass ratio \\
$i$  & degrees & $\mathcal{U}(0, 90)$ & $73^{+1.8}_{-1.5}$ & Binary inclination\\
$M_2$ & \msun & -- & $1.24_{-0.03}^{+0.03}$ & Mass of the unseen compact object\\
$f$  & $\equiv R_1$/$R_\mathrm{L1}$ & -- & $0.90\pm0.02$ & Roche-lobe Filling factor \\
\hline
\end{tabular}
\caption{\label{tab:para} Stellar and binary parameters for J1123. 
$\mathcal{U}(x_{1}, x_{2})$ refers to a uniform distribution 
at range ($x_{1}$, $x_{2}$); 
$\mathcal{N}(x_{0}, \sigma)$ refers to a normal distribution 
with a standard deviation $\sigma$ and centered at $x_{0}$. 
The subscript 1 stands for the visible M dwarf, while 2 stands for the hidden, invisible compact object. All uncertainties represent the one-sigma confidence intervals.
(*) we adopt the surface gravity $\log{g} = 4.67$ reported by the StarHorse$^{[98]}$ catalog, with an augmented uncertainty of $\pm 0.10$ dex as the prior distribution for SED fitting.}
\end{table}

\clearpage

\begin{landscape}
\begin{table}
\renewcommand{\arraystretch}{1.05}	
\centering
  \caption{Observation Log for J1123}
  \centering
  \scriptsize
  \begin{tabular}{@{}lllllllllll@{}}
  \hline
  \multicolumn{1}{c}{\textbf{Facility}} &
  \multicolumn{1}{c}{\textbf{UT shut}} &	  
  \multicolumn{1}{c}{\bf{$\rm HJD_{mid}$}} &
  \multicolumn{1}{c}{\textbf{airmass}} &
  \multicolumn{1}{c}{\textbf{phase}} &
  \multicolumn{4}{c}{\textbf{RV ($\mathrm{km~s^{-1}}$)}} &	  
  \multicolumn{1}{c}{\textbf{HC}} & 
  \multicolumn{1}{c}{\bf{log$(L_\mathrm{H\alpha}/L_\mathrm{bol})$}}\\
  \cline{6-9}
	 & yyyy-mm-dd hh:mm:ss &  &  &  &
  \multicolumn{1}{c}{\texttt{PyHammer}} &
  \multicolumn{1}{c}{H$\alpha$} &
  \multicolumn{1}{c}{H$\beta$}  &
  \multicolumn{1}{c}{H$\gamma$} & $\mathrm{km~s^{-1}}$ & \\	  
  \multicolumn{1}{c}{(1)}&
  \multicolumn{1}{c}{(2)}& 
  \multicolumn{1}{c}{(3)}& 
  \multicolumn{1}{c}{(4)}& 
  \multicolumn{1}{c}{(5)}& 
  \multicolumn{1}{c}{(6)}& 
  \multicolumn{1}{c}{(7)}& 
  \multicolumn{1}{c}{(8)}& 
  \multicolumn{1}{c}{(9)}& 
  \multicolumn{1}{c}{(10)}& 
  \multicolumn{1}{c}{(11)}\\
  \hline
  \multirow{3}{*}{LAMOST}&2015-02-22 17:10:00 
   & 2457076.23049 &	- & 0.931 &
   117.8$_{-3.9}^{+5.4}$  &  98$\pm$6 & 120$\pm$10 & 142$\pm$16 & 0.52 & -4.41\\
   &2015-02-22 17:46:00 & 2457076.25549 &	- & 0.022 
   & -37.5$_{-2.7}^{+2.6}$  & -28$\pm$7 & -21$\pm$8 & -11$\pm$22 & 0.47  & -4.35\\
   &2015-02-22 18:18:59 & 2457076.27841 &	- & 0.106 
   & -160.8$_{-3.5}^{+3.5}$ & -132$\pm$5  & -191$\pm$9 & -201$\pm$9 & 0.42 & -4.45\\
  \hline
  \multirow{8}{*}{P200 DBSP}  
  &2019-03-14 04:38:12 & 2458556.70471 & 1.263 & 0.369 &
  -192.2$_{-3.6}^{+3.6}$
  & -176$\pm$7  & -174$\pm$12 & -136$\pm$24  &-7.66 & -4.15  \\
  &2019-03-14 04:58:34&	2458556.71885 & 1.203 & 0.420 &
  -131.0$_{-1.7}^{+1.7}$ 
  & -117$\pm$6 &	-109$\pm$12 & -90$\pm$11 &-7.68 & -4.05 \\
  &2019-03-14 05:22:18&	2458556.73533 & 1.146 & 0.481 & 
  -42.8$_{-1.5}^{+1.5}$	 
  & -55$\pm$5  & -43$\pm$8  & -1$\pm$13  &-7.71 & -4.02  \\
  &2019-03-14 05:42:40&	2458556.74947 & 1.106 & 0.532 & 
  37.5$_{-1.5}^{+1.4}$	
  & 19$\pm$5 & 42$\pm$7 & 67$\pm$14  &-7.74 & -3.97 \\
  &2019-03-14 06:13:40&	2458556.77100 & 1.061 & 0.611 & 
  158.5$_{-1.4}^{+1.4}$
  &	134$\pm$5 &	150$\pm$8 & 187$\pm$12  &-7.79 & -4.09 \\
  &2019-03-14 06:34:02&	2458556.78514 & 1.039 & 0.662 &	
  210.8$_{-1.5}^{+1.4}$	 
  &	189$\pm$6 & 211$\pm$9 & 234$\pm$15  &-7.82 & -4.16 \\
  &2019-03-14 06:54:23&	2458556.79928 & 1.023 & 0.714 & 
  241.6$_{-2.2}^{+1.2}$	 
  &	209$\pm$6  &	219$\pm$9  & 259$\pm$10  &-7.85 & -4.04 \\
  &2019-03-14 07:14:45&	2458556.81342 & 1.012 & 0.766 &	
  250.2$_{-1.8}^{+2.0}$
  & 216$\pm$5  & 225$\pm$11 & 256$\pm$19  &-7.88 & -4.03 \\ 
  \cline{2-11}
	&2020-06-24 04:20:35 & 2459024.68555 & 1.307 & 0.354 & 
	-210.3$_{-2.1}^{+2.2}$ 
	& -232$\pm$4  & -227$\pm$8  & -195$\pm$9 & -21.82 & -3.93 \\
	&2020-06-24 04:40:57 & 2459024.69969 & 1.389 & 0.406 & 
	-150.7$_{-7.4}^{+3.7}$
	& -172$\pm$4  & -164$\pm$5  & -158$\pm$13 & -21.83 & -3.94\\
  \hline  
  \end{tabular}\label{tab:log}
  \vspace{1ex}

  {\raggedright Notes: column(1): facility;
 column(2): the UTC shutter open time;
 column(3): the Heliocentric Julian Date in the mid-time of each exposure;
 column(4): the airmass recorded at the UTC shut;
 column(5): the orbital phase;
 column(6): RVs reported by the \texttt{PyHammer};
 column(7)-(9): RVs of H$\alpha$, H$\beta$, and H$\gamma$, respectively; 
 column(10): Heliocentric correction. 
 All reported RVs in column(6)-(9) are measured in the Heliocentric rest-frame.
 Namely, all spectra have been corrected by the HC before the RVs to be measured;
 column(11): normalised H$\alpha$ luminosity. 
 \par} 
\end{table}

\end{landscape}

\begin{table}
\renewcommand{\arraystretch}{1.05}	
\centering
  \caption{Broadband SED photometry}
  \centering
  \footnotesize
	\begin{tabular}{@{}llllll@{}}
	\hline
	Survey & Band & $N_\mathrm{obs}$ & AB mag & Vega mag & $A_{\lambda}$ (mag) \\
	(1)    & (2)  & (3)              & (4)    & (5)      & (6)              	\\	
	\hline
	GALEX$^{[43]}$
		 & 	FUV	 &	1 & 	-	 & 	-	 & 	-	 \\
		 & 	NUV$^{*}$	 &	1 & 	22.731$\pm$0.516	 & 	-	 & 	0.160$\pm$0.008	 \\ 
	\hline
	SDSS$^{[44]}$
		 & 	u$^{*}$	 &	1 & 	19.724$\pm$0.040	 & 	-	 & 	0.091$\pm$0.005	 \\ 
		 & 	g	 &	1 & 	17.318$\pm$0.005	 & 	-	 & 	0.071$\pm$0.004	 \\ 
		 & 	r	 &	1 & 	15.954$\pm$0.003	 & 	-	 & 	0.051$\pm$0.003	 \\ 
		 & 	i	 &	1 & 	15.167$\pm$0.003	 & 	-	 & 	0.039$\pm$0.002	 \\ 
		 & 	z	 &	1 & 	14.756$\pm$0.004	 & 	-	 & 	0.028$\pm$0.001	 \\ 
	\hline
	\textit{Gaia} EDR3$^{[18]}$
		 & 	GP	 &	320 & 	-	 & 	15.832$\pm$0.005		 & 	0.054$\pm$0.003	 \\ 
		 & 	BP	 &	36 &	- & 16.845$\pm$0.016		& 	0.064$\pm$0.003	 \\ 
		 & 	RP	 &	37 & - & 14.810$\pm$0.011		& 	0.037$\pm$0.002	 \\ 		 
	\hline
	ZTF$^{[34]}$
		 & g	& 827 & 	17.292$\pm$0.0007 & 	-	& 0.070$\pm$0.004	 \\ 
		 & r	& 899 & 	15.885$\pm$0.0005 & 	-	& 0.049$\pm$0.003	 \\ 
		 & i	& 66 & 	15.069$\pm$0.0018 & 	-	& 0.036$\pm$0.002	 \\ 		 
	\hline	
	PanSTARRS$^{[45]}$ 
		 & 	g	 &	12 & 	17.149$\pm$0.016	 & 	-	 & 	0.069$\pm$0.004	 \\ 
		 & 	r	 &	15 & 	15.971$\pm$0.006	 & 	-	 & 	0.051$\pm$0.003	 \\ 
		 & 	i	 &	28 & 	15.244$\pm$0.012	 & 	-	 & 	0.039$\pm$0.002	 \\ 
		 & 	z	 &	13 & 	14.835$\pm$0.016	 & 	-	 & 	0.030$\pm$0.002	 \\ 
		 & 	y	 &	16 & 	14.628$\pm$0.014	 & 	-	 & 	0.025$\pm$0.001	 \\ 
	\hline
	2MASS$^{[46]}$
		 & 	J	 &	1 & 	-	 & 	13.536$\pm$0.024	 & 	0.017$\pm$0.001	 \\ 
		 & 	H	 &	1 & 	-	 & 	12.878$\pm$0.026	 & 	0.011$\pm$0.001	 \\ 
		 & 	Ks	 &	1 & 	-	 & 	12.699$\pm$0.023	 & 	0.007$\pm$0.000	 \\ 
	\hline
	AllWISE$^{[47,48]}$
		 & 	W1	 &	30 & 	-	 & 	12.554$\pm$0.023	 & 	0	 \\ 
		 & 	W2	 &	30 & 	-	 & 	12.505$\pm$0.024	 & 	0	 \\ 
		 & 	W3	 &	15 & 	-	 & 	$>$12.385	 & 	0	 \\ 
		 & 	W4	 &	16 & 	-	 & 	$>$9.081	 & 	0	 \\ 
	\hline
  \end{tabular}\label{tab:SED}
  
  \vspace{1ex}
  {\raggedright Notes: 
  column (1): survey name; 
  column (2): filter name;
  column (3): number of observations conducted by each band;
  columns (4) and (5): magnitude and the corresponding statistical (random) uncertainty;  
  column (6): extinction at each band's 
  effective wavelength $\lambda_\mathrm{eff}$
  (calculated by applying the R$_\mathrm{V}$ $=$ 3.1 
  reddening law$^{[52]}$
  and the Cardelli extinction curve$^{[51]}$, 
  given that $E(B-V)$ $=$ 0.019 $\pm$ 0.001 mag. 
  The extinction for WISE W1-W4 bands are assumed to be zero);
  (*) GALEX NUV and SDSS u-band were not used to the SED fitting.
 \par} 
\end{table}

\begin{table}
\renewcommand{\arraystretch}{1.1}	
	\centering
	\footnotesize
  \caption{Stellar parameters of AD Leo and AU Mic}
  \centering
	\begin{tabular}{@{}llllll@{}}
	\hline
	\textbf{Source} & \textbf{Spectral type} & Distance & Radius & \textbf{$T_\mathrm{eff}$} 
	& \textbf{$\mathrm{\log}(L_\mathrm{H\alpha}/L_\mathrm{bol})$} \\
	& & $\mathrm{pc}$ & $R_{\odot}$ & $\mathrm{K}$ & \\
	\hline
	AD Leo & dM3e & 4.9 & 0.37 & 3350 & -3.51 \\
	(reference) & (1)(2)(3) & (4)(2)(3) & (2) & (5)(2) & (6)(7) \\
	\hline	
	AU Mic & dM1 & 9.72 & 0.75 & 3700 & -3.62 \\
	(reference) & (8)(9) & (9) & (10)(9) & (11) & (6)$^{(*)}$\\
	\hline
  \end{tabular}\label{tab:ADAU}
  \vspace{1ex}

  {\raggedright Notes: (1) Henry et al. 1994$^{[99]}$
  (2) Favata et al. 2000$^{[100]}$
  (3) Walkowicz et al. 2008$^{[101]}$
  (4) Reid et al. 1995$^{[102]}$
  (5) Jones et al. 1996$^{[103]}$
  (6) Rugheimer et al. 2015$^{[84]}$
  (7) Walkowicz et al. 2009$^{[104]}$
  (8) Torres et al. 2006$^{[105]}$
  (9) Plavchan et al. 2020$^{[106]}$
  (10) White et al.2015$^{[107]}$
  (11) Plavchan et al. 2009$^{[108]}$
  (*) evaluated by using the Equation (1) in Rugheimer et al. 2015$^{[84]}$
 \par} 
\end{table}

\renewcommand{\figurename}{Supplementary Figure}
\renewcommand\thefigure{\arabic{figure}}    
\setcounter{figure}{0}

\newpage
\begin{landscape}
\centering
\begin{figure}
\includegraphics[width=1.0\columnwidth]{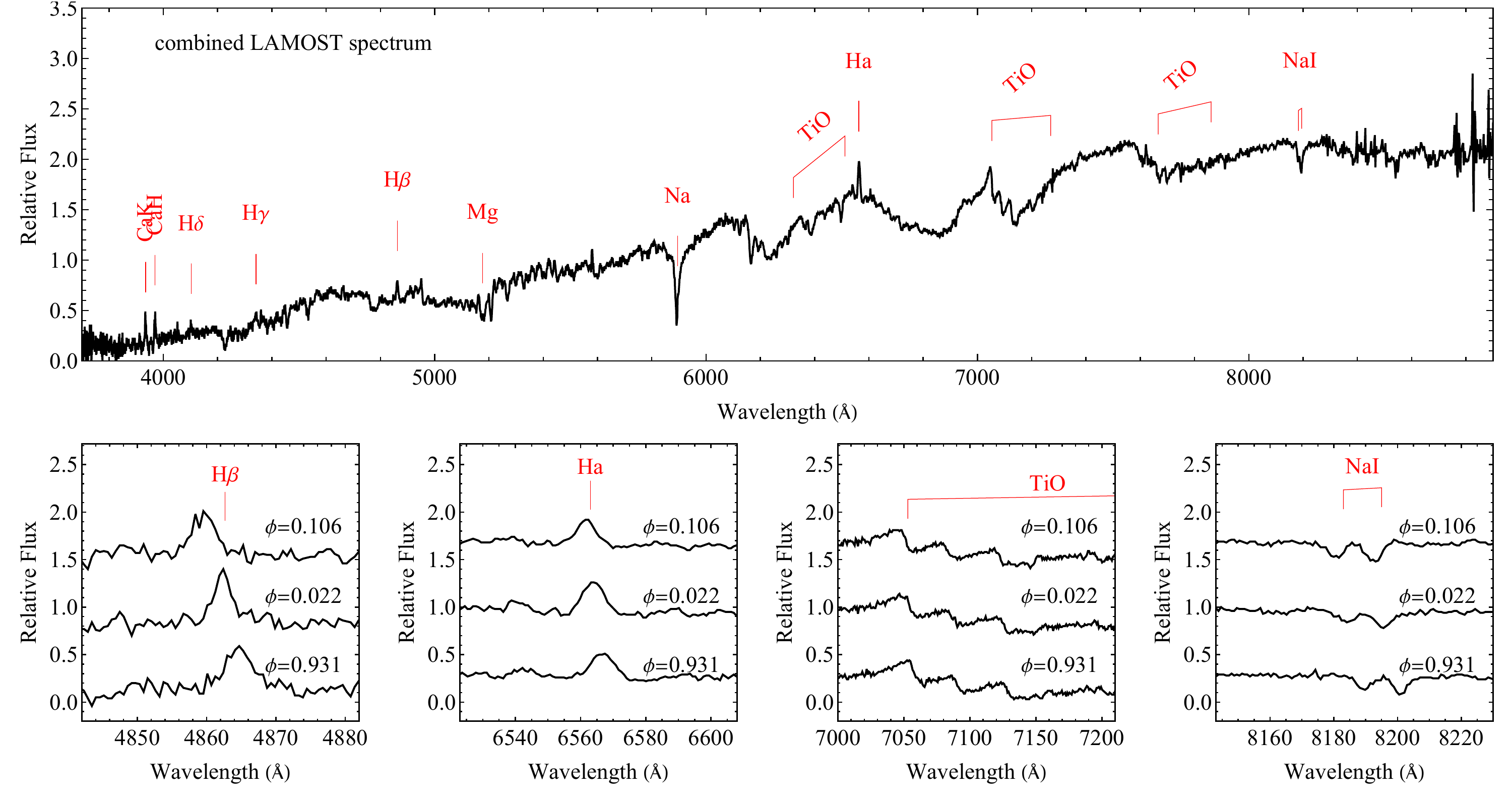}
\caption{The LAMOST low-resolution spectrum for J1123.
Upper panel: the combined spectrum of LAMOST.
Lower panels: zoom-in in the vicinity of 
${\rm H\beta}$~4863\AA{}, ${\rm H\alpha}$~6564\AA{}, 
${\rm TiO}$~7050-7200\AA{}, and ${\rm Na I }$~8200\AA{} (doublet),
for three consecutive exposures taken by LAMOST on Feb 22, 2015.
}
\label{fig:speclamo}
\end{figure}
\end{landscape}
\clearpage

\newpage
\begin{landscape}
\centering
\begin{figure}
\includegraphics[width=1.0\columnwidth]{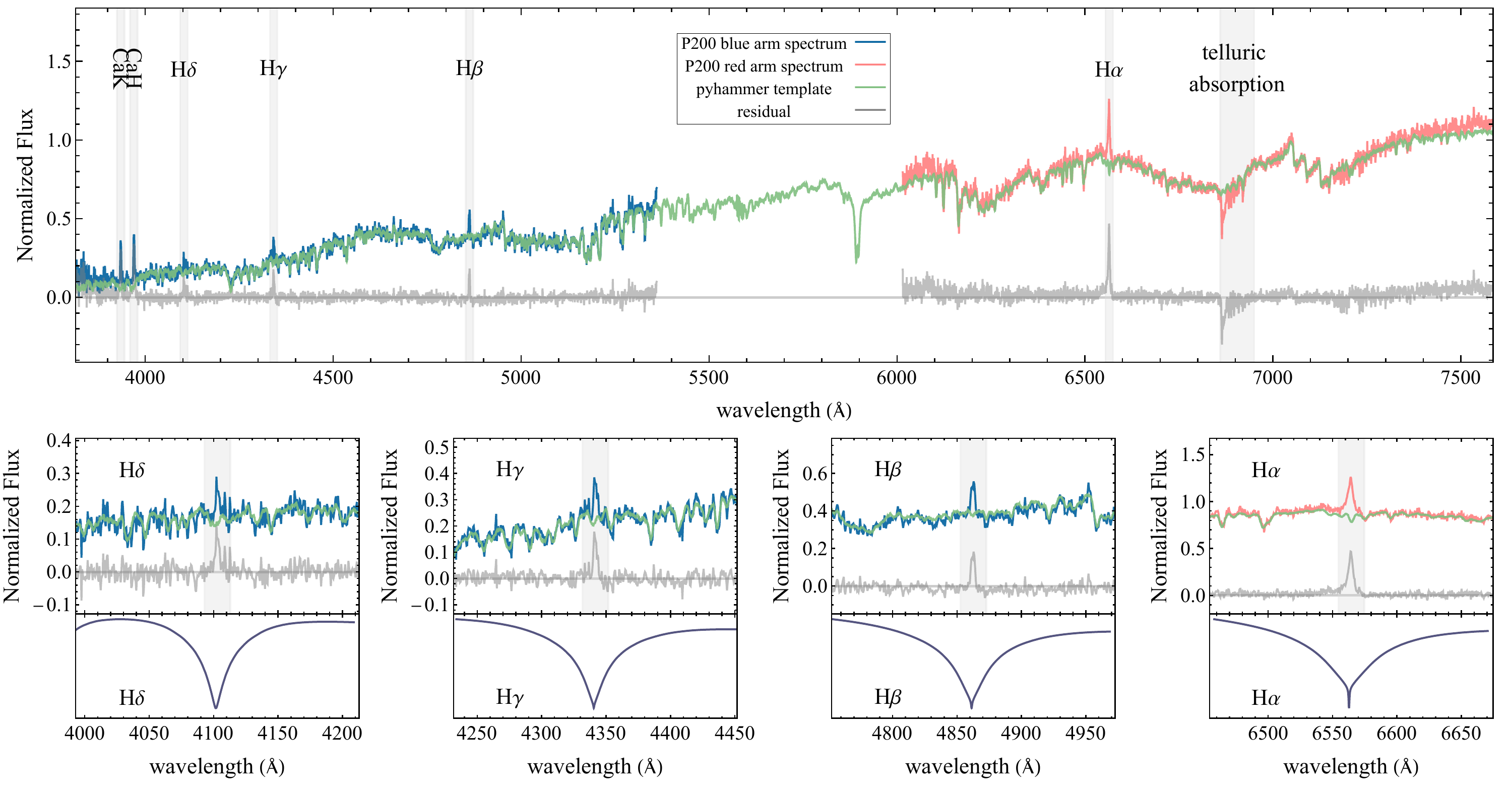}
\caption{The P200 spectrum for J1123.
Upper panel: the P200 spectrum (the eighth exposure taken 
by DBSP on Mar 14, 2019; the mid-time of the exposure corresponds to the orbital 
phase $\phi$ $=$ 0.766) and the best-fit template. 
The spectrum is shifted to the rest-frame in vacuum.
The best-fitted \texttt{PyHammer} template (green)
has a spectral type $=$ M1 and metalicity $Z$ $=$ 0.
Gray curve is the residual ($=$ observation $-$ template).  
Telluric absorption, Balmer emission lines, 
and \mbox{Ca II H\&K} emission lines are shaded by shallow gray regions.
Lower panels: zoom-in in the vicinity of 
${\rm H\alpha}$, ${\rm H\beta}$, ${\rm H\gamma}$, and ${\rm H\delta}$.
In the bottom sub-panels, the corresponding broad Balmer absorption lines of a theoretical WD spectrum$^{[93]}$ (purple curve)
with effective temperature $=$ 10000 K are shown for comparison. 
It is evident that no broad absorption feature of a WD is detected in our observations, as suggested by the residuals.
}
\label{fig:specP200}
\end{figure}
\end{landscape}
\clearpage

\begin{figure}
\centering
\includegraphics[width=0.618\linewidth]{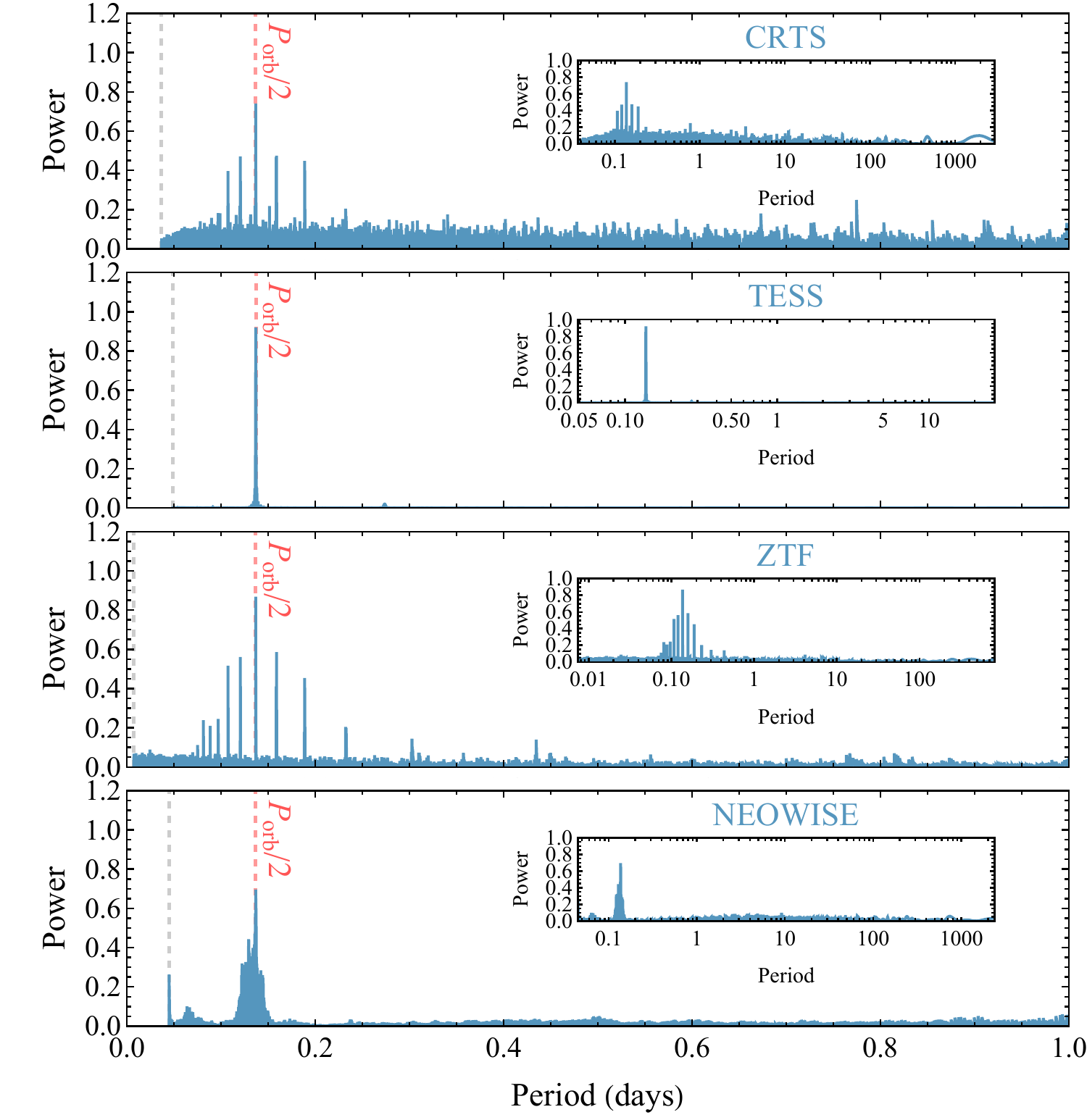}
\caption{The Lomb-Scargle powers of CRTS, TESS, ZTF r-band, 
and NEOWISE W1-band light curves (from top to bottom, respectively). 
The highest power (marked by the vertical red dashed lines) 
reports half the true orbital period of the ellipsoidal light curves.
Vertical gray dashed lines mark the highest frequency of the period search grid,
corresponding to a Nyquist factor $=$ 500 for CRTS, ZTF, and NEOWISE 
and a Nyquist factor $=$ 1 for TESS.
The inset panel shows the power within the full searched period (frequency) window.} 
\label{fig:LSpower}
\end{figure}

\begin{figure}
\centering
\includegraphics[width=1.0\columnwidth]{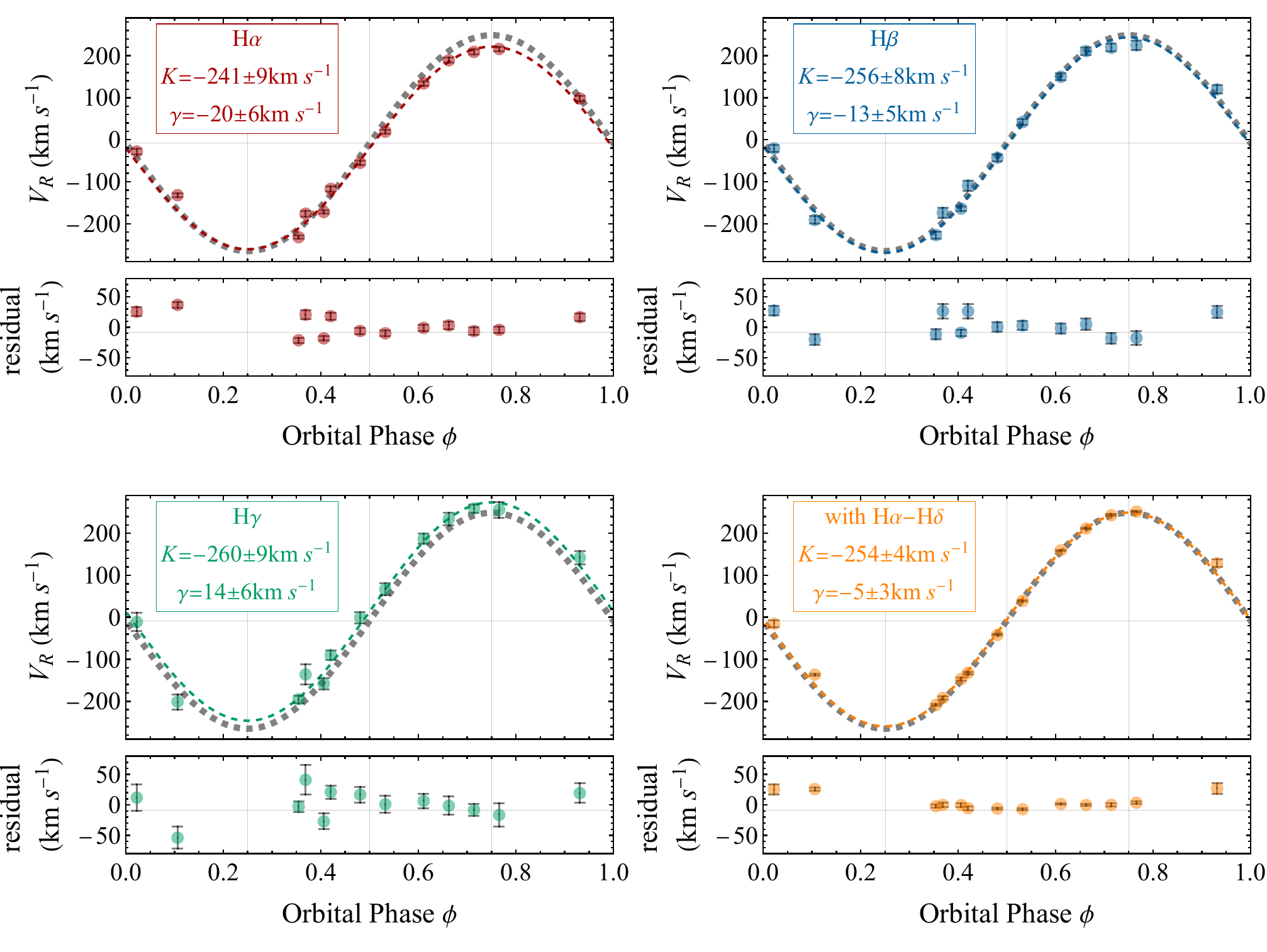}	
\caption{The RV curves by fitting the
H$\alpha$ (red), H$\beta$ (blue), H$\gamma$ (green) lines,
and by fitting the full spectrum (orange) with all Balmer lines included. 
Data points refer to Supplementary Table~\ref{tab:log} 
and error bars represent the $1$-sigma uncertainties.
As for comparison, the gray dashed line in each panel
is the RV curve obtained by fitting the full spectrum 
with all Balmer lines excluded 
(i.e., by fitting only the photospheric absorption lines of the M star).
}
\label{fig:rvs}
\end{figure}

\begin{figure}
\centering
\includegraphics[width=1.0\linewidth]{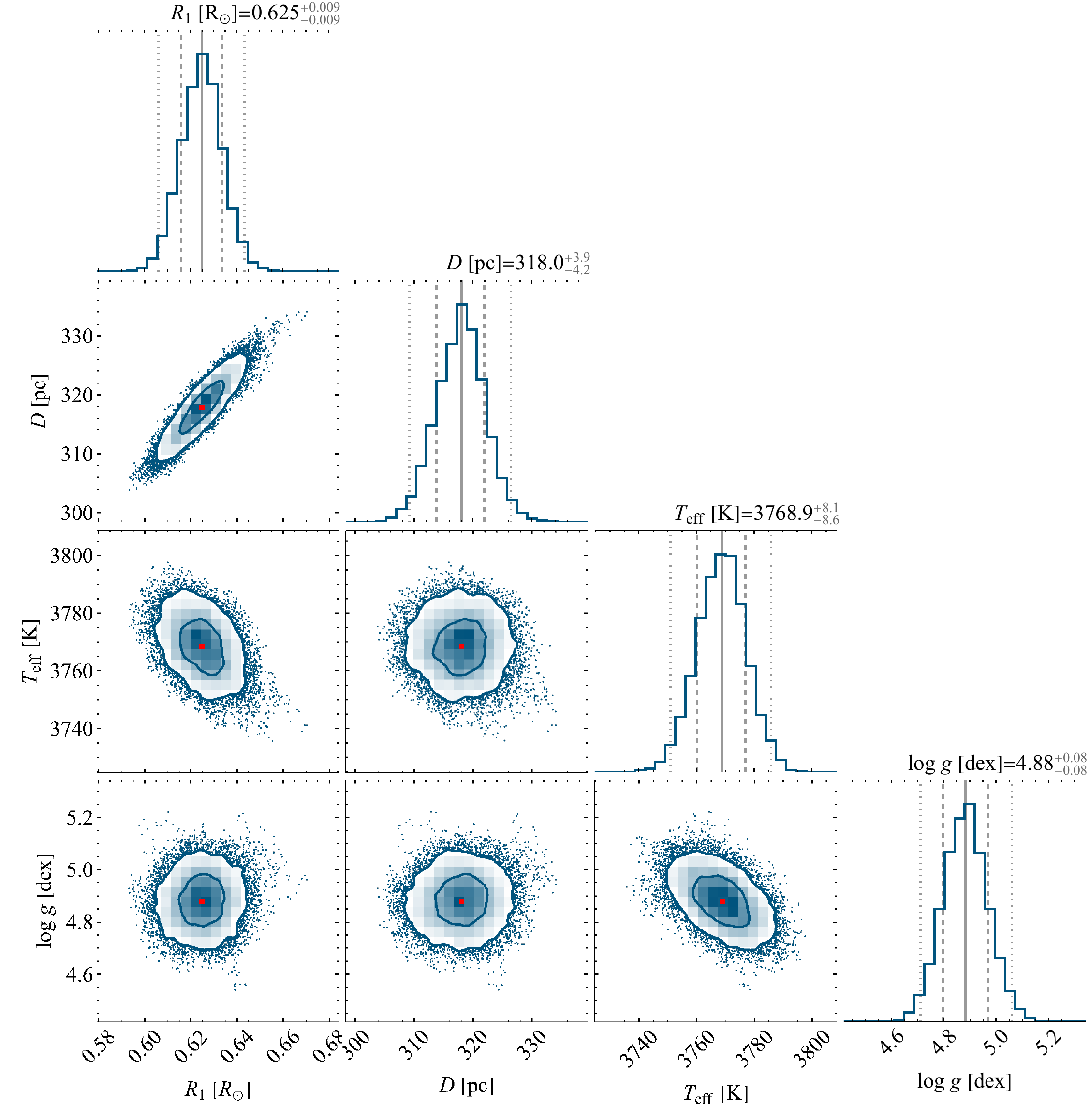}
\caption{The corner plot for the broadband SED fitting. 
The histograms indicate the marginal distributions for the fitted stellar parameters and the density plots indicate the joint distributions.
In the histograms, the vertical dashed lines and the dotted lines represent the 1-sigma and the 2-sigma uncertainties, respectively.} 
\label{fig:sed_corner}
\end{figure}

\begin{figure}[ht]
\centering
\includegraphics[width=1.0\linewidth]{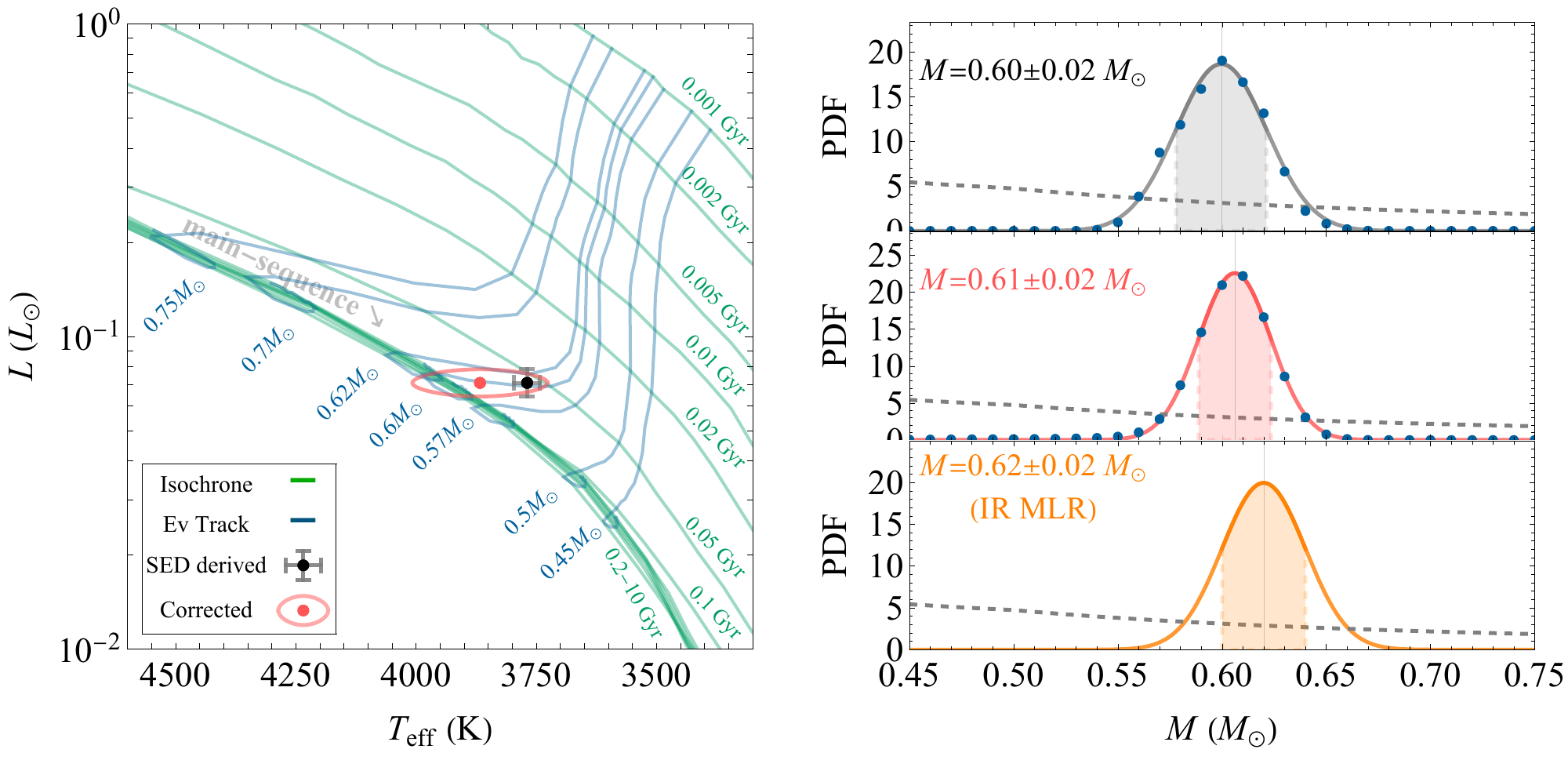}
\caption{The stellar properties of the visible M dwarf. 
Left panel: the position of J1123 in the HR diagram. 
Blue lines are the BT-Settl evolutionary tracks for stellar masses 
$M$ $=$ 0.45, 0.50, 0.57, 0.60, 0.62, 0.70, and 0.75$M_{\odot}$.
Green lines are the isochrones at selected
ages $=$ 0.001, 0.002, 0.005, 0.01, 0.02, 0.05, 
0.1, 0.2, 0.5, 1, 2, 5, and 10 Gyrs. 
Note that the isochrones of 0.1-10 Gyrs almost lie on top of each other
since M dwarfs do not move much once entering the main-sequence.
The black dot is the SED derived (uncorrected) 
position of J1123 in the HR diagram, 
with error bars being the three-sigma uncertainties. 
The pink dot and ellipse are the corrected position and uncertainty
by taking into account the influence of the tidal distortion effect.
Right panels: the posterior distributions of the M dwarf's mass,
by using the uncorrected (SED derived; right upper panel) 
and the corrected $T_\mathrm{eff}$ (right middle panel), respectively.
The posterior distributions (blue dots) are calculated by fitting the 
isochrones in 0.01 $M_{\odot}$ mass bins and fitted with a Gaussian 
(black and pink curves). The gray dashed line in each panel is the 
Kroupa initial mass function$^{[55]}$, which is used as the 
prior. Also shown in the right lower panel is the mass inferred from 
the IR (2MASS Ks band) MLR$^{[60]}$.} 
\label{fig:m1}
\end{figure}

\newpage
\begin{landscape}
\begin{figure}
\centering
\includegraphics[width=1.0\linewidth]{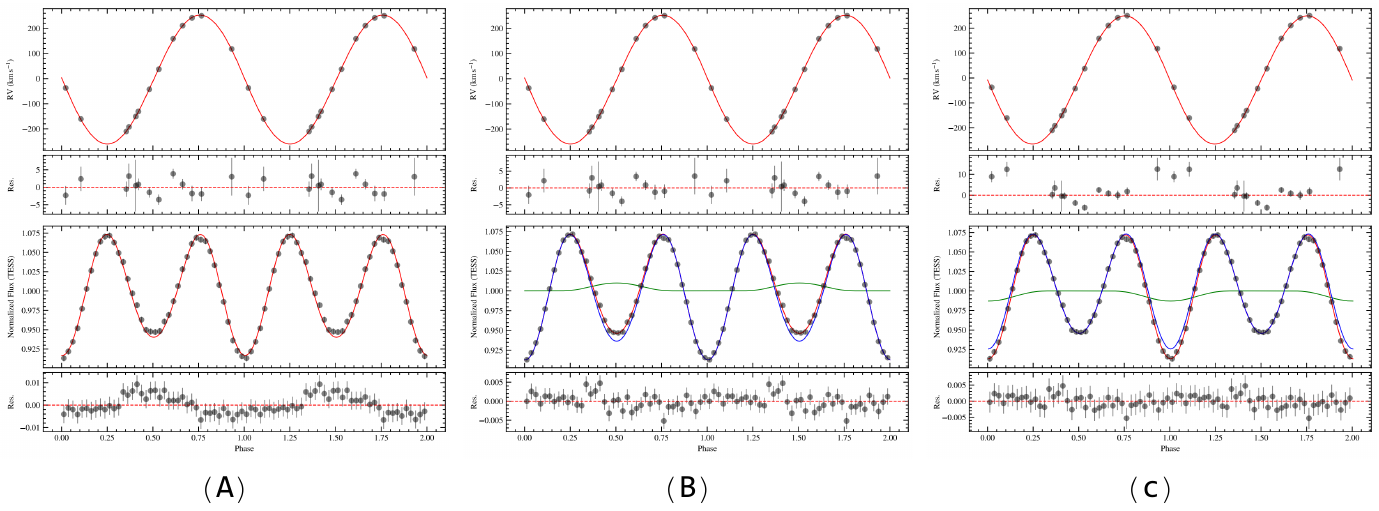}
\caption{\texttt{PHOEBE}'s best-fitting results.
Upper panels: 
the best-fitting radial velocities 
(the red curve) versus the observations (the black dots) and the residuals 
(i.e., the differences between the fit and the data). 
Lower panels: 
the best-fitting flux variations (the red curve) versus the \texttt{TESS} 
observations (the black dots) and the residuals. 
The error bars correspond to the $1\sigma$ uncertainties. 
For model A (column A), 
we only consider the ellipsoidal modulations.
It is evident that the light curve residuals show periodic variations. 
For model B (column B), 
The model light curve (the red curve) is a superposition of the 
ellipsoidal modulations (the blue curve) and a hotspot (the green 
curve).
For model C (column C), 
the model light curve (the red curve) is superposition of the 
ellipsoidal modulations (the blue curve) and a coldspot (the green 
curve). } 
\label{fig:model-abc}
\end{figure}
\end{landscape}
\clearpage

\begin{figure}
\centering
\includegraphics[width=0.9\columnwidth]{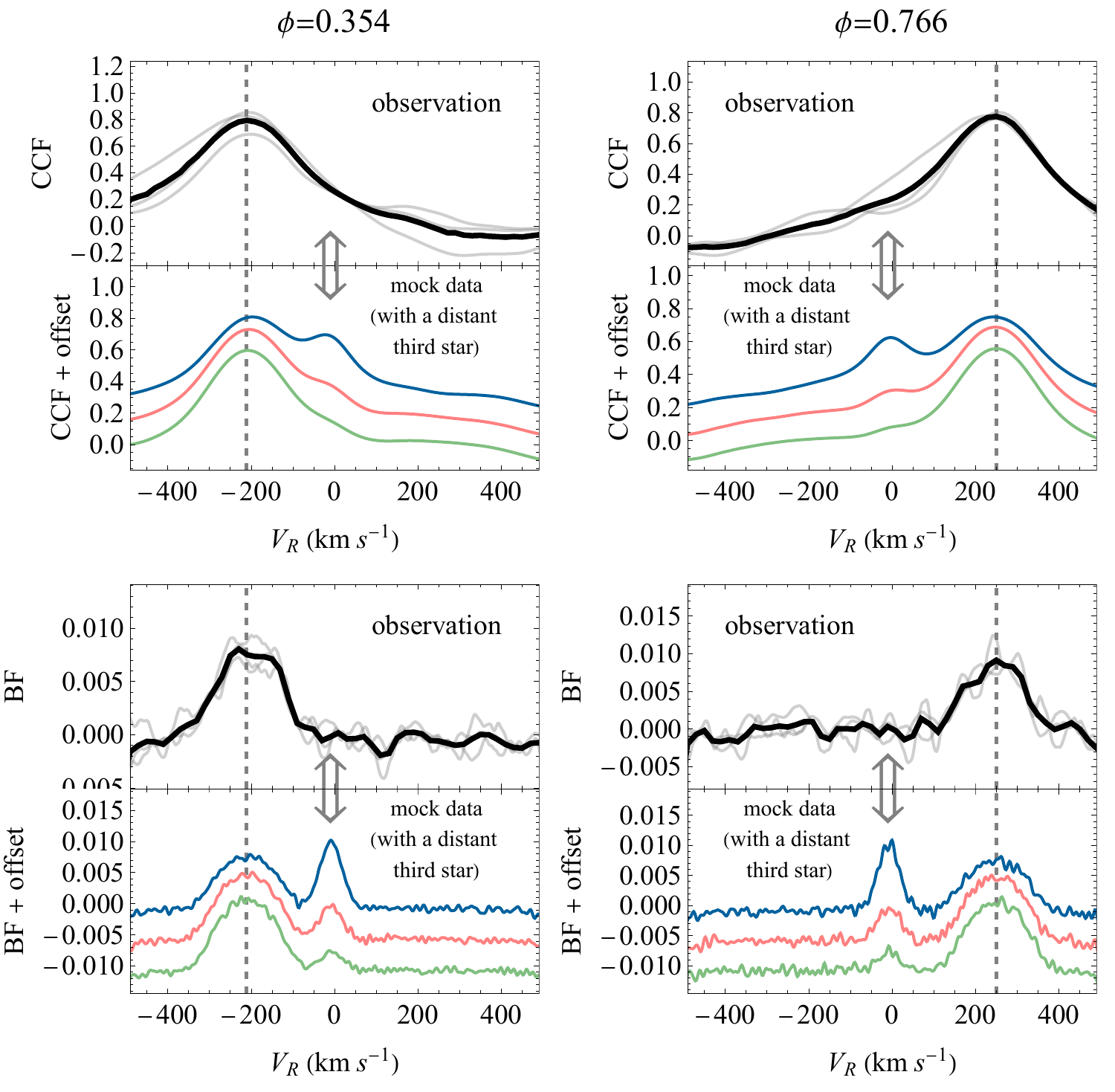}	
\caption{
The CCFs (top row) and the BFs (bottom row) for two P200 spectra at phases 0.354 
(left column) and 0.766 (right column). Their corresponding RVs are indicated by the vertical 
dashed lines. 
For each panel, the upper sub-panel is the result of our P200 observations (black curve) obtained by 
averaging three shallow gray curves calculated using three different wavelength ranges 
(see Methods); the lower sub-panel is the result of the corresponding mock data in the case of a triple system. The blue, pink, and green curves represent the results for the case where the  
luminosity of the third star is 0.5, 0.2, and 0.1 times the luminosity of our visible M star, respectively 
(small vertical offsets are added for clarity). For the mock data, 
there is a prominent secondary peak from the third object 
in the CCF or the BF; such peak is absent in observations (indicated by the gray arrows). 
Thus a distant third object with luminosity $\gtrsim$ 0.1 times the luminosity of the 
M star can be safely ruled out.
}
\label{fig:ccfbf}
\end{figure}

\newpage
\begin{landscape}
\begin{figure}
\centering
\includegraphics[width=1.0\linewidth]{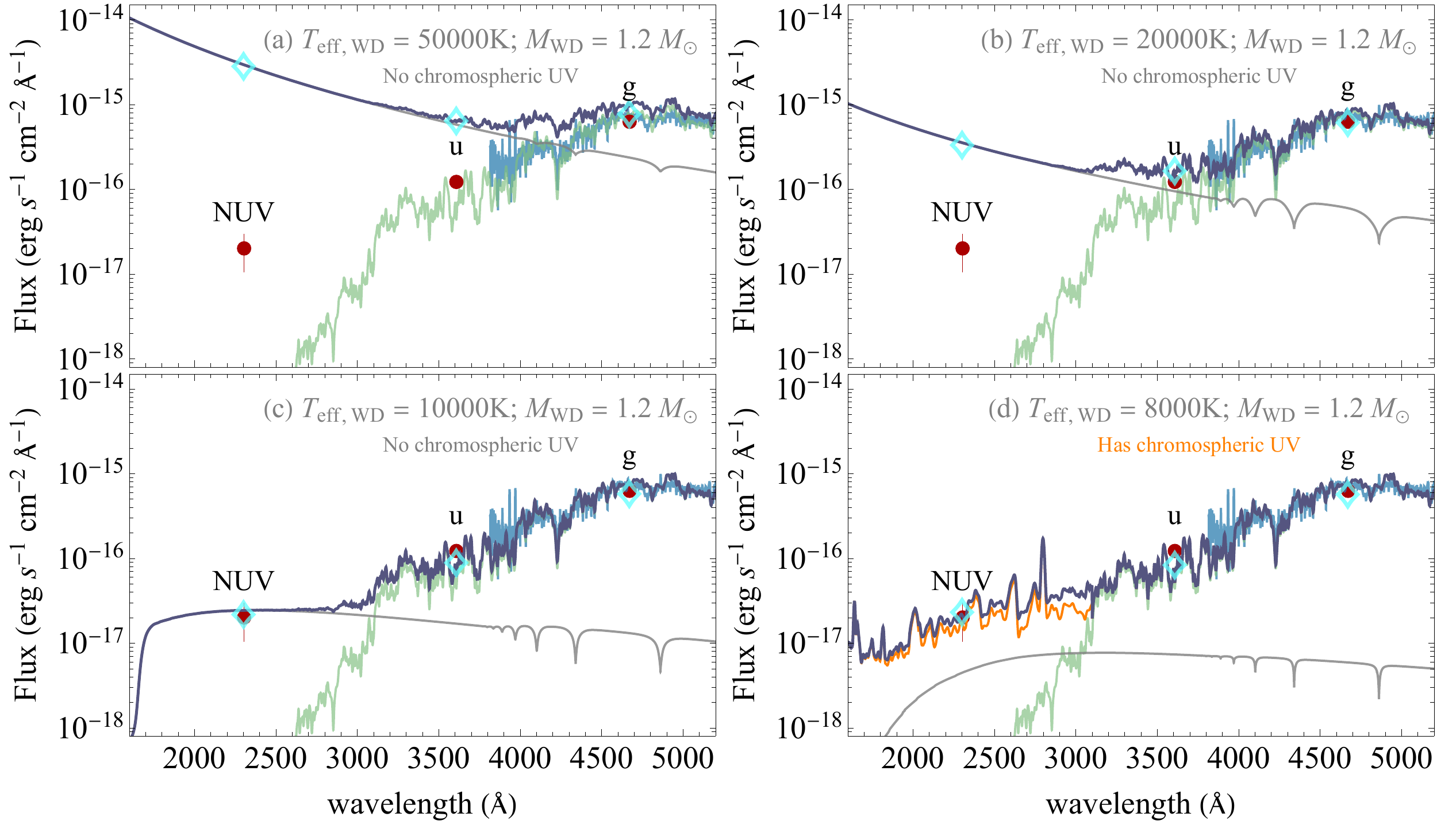}
\caption{
Comparison of the P200 blue arm spectrum (light blue curve) with the composite SED (purple) 
of the M dwarf (green) plus a WD (gray).
The observed photometry (red dots) and the synthetic photometry of the composite SED (cyan dimonds) are also shown for comparison. 
By using the known distance 318 pc,
we scale on SED with four DA-type WD templates 
with the same mass ($M_\mathrm{WD} = 1.2\ M_{\odot}$) 
and the same surface gravity ($\log{g}$ $=9.0$), 
but different temperatures (i.e., 
$T_\mathrm{eff, WD}=5\times 10^4$ K, $2\times 10^4$ K, $10^4$ K, and 
$8\times 10^3$ K).
Only WDs with $T_\mathrm{eff, WD}$ $\lesssim 10^4$ K can well explain 
the observations. Panel (d) shows the realistic case, 
in which the composite SED also contains a substantial amount of 
chromospheric UV emission (orange curve; taken as the scaled mean 
spectrum of AD Leo and AU Mic for demonstration).  
}
\label{fig:sedwd}
\end{figure}
\end{landscape}
\clearpage

\begin{figure}
\centering
\includegraphics[width=1.0\linewidth]{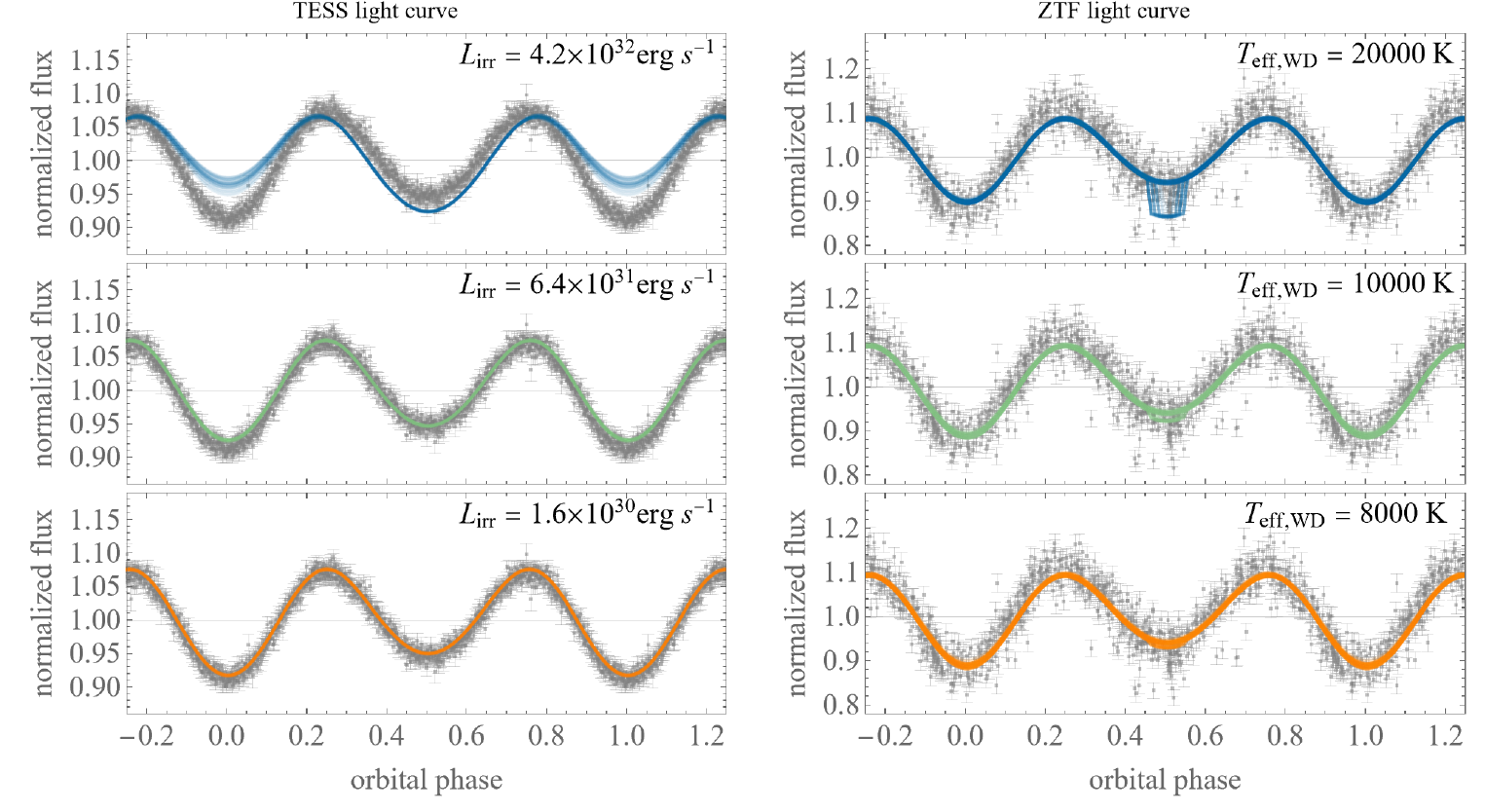}
\caption{Comparison of 
the \texttt{PHOEBE} simulated TESS light curves (left panels; coloured lines) 
and ZTF light curves (right panels; coloured lines) with the observations. 
Gray points are the observational data 
and the error bars represent the 1-sigma uncertainties.
We use 100 sets of orbital 
parameters randomly drew from the distributions of the \texttt{PHOEBE}'s 
orbital solution to simulate each case described below. 
For the NS (left panels), 
three different irradiation luminosities: 
$4.2 \times 10^{32}\ \mathrm{erg~s^{-1}}$, 
$6.4 \times 10^{31}\ \mathrm{erg~s^{-1}}$, and 
$1.6 \times 10^{30}\ \mathrm{erg~s^{-1}}$ are investigated.
It is evident that for a NS with 
$L_{\rm irr} \gtrsim 6.4 \times 10^{32}~\mathrm{erg~s^{-1}}$,
the irradiation effect showing on the light curve is inconsistent with the \texttt{TESS} observations. 
Therefore we constrain the irradiation luminosity to be less than 
$6.4 \times 10^{32}~\mathrm{erg~s^{-1}}$.
For the WD (right panels), three different effective temperatures: 
20000 K, 10000 K, and 8000 K are investigated.
For a hot WD with $T_{\rm eff,~WD}$ = 20000 K,
when the orbital inclination angle is large (nearly edge-on case), a prominent eclipse 
is expected at the M star's inferior conjunction phase $\phi = 0.5$; otherwise, there are no eclipses. 
For a WD with $T_{\rm eff,~WD} \lesssim 10000$ K, 
possible eclipse becomes hard to be identified given the photometric precision of our light curve. The simulation 
suggests that if the compact object in J1123 is a WD, the WD must be cold and 
massive.
} 
\label{fig:ns}
\end{figure}

\clearpage

\end{document}